\documentclass[11pt]{article}
\usepackage{amsmath,amssymb,amsthm}
\usepackage{graphics,epsfig}
\usepackage{hyperref}
\usepackage{natbib}
\usepackage{color}
\usepackage{graphicx}
\usepackage{caption}
\usepackage{subcaption}
\usepackage{float}
\usepackage{dsfont}
\usepackage{mathrsfs}
\usepackage{multirow}
\usepackage{comment}
\usepackage{setspace}
\usepackage{bbm}
\usepackage{bm}
\usepackage{amsfonts}
\usepackage{xcolor}
\usepackage{pslatex}
\usepackage{setspace}
\def \ra {\rightarrow}

\def \oo {\infty}
\def \E {\mathbb{E}}

\def \a {\alpha}
\def \be {\beta}

\newtheorem{definition}{\bf Definition}

	\newtheorem{remark}{\bf Remark}

	\newtheorem{theorem}{\bf Theorem}
	
	\newtheorem{prop}{\bf Proposition}
	\newtheorem{lem}[theorem]{\bf Lemma}
	\newtheorem{as}{\bf Assumption}


\newcommand{\pkg}[1]{{\fontseries{b}\selectfont #1}}   % for package name

\begin{document}
	
	\title{\bfseries Bayes Factor of Zero Inflated Models under Jeffereys Prior}
	
	\author{
		Paramahansa Pramanik \footnote{e-mail: {\small\texttt{ppramanik@southalabama.edu}} }\; \footnote{Department of Mathematics and Statistics, University of South Alabama, 411 North University Boulevard, Mobile, AL 36688, USA.} \; \; Arnab Kumar Maity \footnote{email:{\small\texttt{arnab.maity@pfizer.com}}}\; \footnote{Pfizer Inc., 10777 Science Center Drive, San Diego, CA 92121, USA.}}
	
	\date{\today}
	\maketitle
	
\subparagraph{Abstract.}
Microbiome omics data including 16S rRNA reveal intriguing dynamic associations between the human microbiome and various disease states. Drastic changes in microbiota can be associated with factors like diet, hormonal cycles, diseases, and medical interventions. Along with the identification of specific bacteria taxa associated with diseases, recent advancements give evidence that metabolism, genetics, and environmental factors can model these microbial effects. However, the current analytic methods for integrating microbiome data are fully developed to address the main challenges of longitudinal metagenomics data, such as high-dimensionality, intra-sample dependence, and zero-inflation of observed counts. Hence, we propose the Bayes factor approach for model selection based on negative binomial, Poisson, zero-inflated negative binomial, and zero-inflated Poisson models with non-informative Jeffreys prior. We find that both in simulation studies and real data analysis, our Bayes factor remarkably outperform traditional Akaike information criterion and Vuong's test. A new R package BFZINBZIP has been introduced to do simulation study and real data analysis to facilitate Bayesian model selection based on the Bayes factor.

\subparagraph{Key words:} Negative binomial distribution; Zero inflated negative binomial distribution, Poisson distribution, zero inflated poisson distribution; Bayes factor; non-informative Jeffreys prior; Microbiome.	

\section{Introduction}

The human microbiome consistes of the collection of estimated $3.0\times 10^{13}$ \citep{sender2016revised} bacteria and $3.3\times 10^6$ genes \citep{qin2010human,jiang2021}. The human microbiome analysis impose a drastic impact on human health and disease \citep{ursell2012defining}. In recent years, microbiome studies have been successfully identified disease-associated bacteria taxa in type 2 diabetes \citep{karlsson2013gut}, liver cirrhosis \citep{qin2014alterations}, inflammatory bowl disease \citep{halfvarson2017dynamics,kakkat2023cardiovascular}, and melanoma patients responsive to cancer immunotherapy \citep{frankel2017metagenomic,jiang2021,dasgupta2023frequent,hertweck2023clinicopathological,khan2023myb,vikramdeo2023profiling}. Quantification of of the human microbiome usually being proceeded by 16s rRNA sequencing or metagenomic shotgun sequencing, where sequence read counts are often summarized into a taxa count table \citep{jiang2023flexible}. Bioinformatics tools like quantitative insights into microbial ecology (QIIME) and mothur are used for analyzing raw 16S rRNA sequencing data \citep{jovel2016characterization,zhang2020}. In this literature the word \emph{taxa} means operational taxonomic units or other taxonomic or functional groups of bacterial sequences \citep{jiang2023flexible, altaweel2022detecting}. Although innovations in sequencing technology continue to prosper in microbiome studies, the statistical methods used in this field fail to catch up with these advanced sequences \citep{jiang2021}. For example, metagenomic shotgun sequencing generates an increasingly large amount of sequence reads which give species or confine level taxonomic resolution \citep{segata2012metagenomic}. The subsequent statistical analysis compares whether a species is linked to a phenotypic state or experimental condition \citep{jiang2021,polansky2021motif, maity2018bias}.

One commonly used statistical approach in microbiome community involves comparing multiple taxa \citep{chen2012associating,kelly2015power,wu2016adaptive,jiang2021}. These approaches do not identify differentially abundant species, which makes clinical interpretation, mechanistic insights, and biological validations difficult \citep{jiang2021,pramanik2023optimization}. An alternative approach is to interrogate each individual bacterial taxa for different groups or conditions. \cite{la2015hypothesis} use Wilcoxon rank sum or Kruskal-Wallis tests for groupwise comparisons on microbiome compositional data. In more recent years, RNA-seq methods have been adapted to microbiome studies, such as the negative-binomial regression model in DESeq2 \citep{love2014moderated} and overdispersed Poisson model in edgeR \citep{robinson2010edger, maity2018salicylic}. However, these approaches are not optimized for microbiome datasets \citep{jiang2021}. Microbial abundance is influenced by covariates like metabolites, antibiotics, and host genetics \citep{pramanik2022lock,pramanik2022stochastic,pramanik2023path}. To account for these confounding variables, the association between microbiome and clinical confounders must be quantified. Pairwise correlations between all taxa and covariates are commonly used, but this method may be underpowered \citep{kinross2011gut,maier2018extensive,zhu2018precision}. Other approaches have been proposed to detect covariate-taxa associations, but these ignore the taxon-outcome associations \citep{pramanik2023optimal,pramanik2023path1}. Recently, \cite{li2018conditional, schweizer2022488p} developed a multivariate zero-inflated logistic normal model to quantify the associations between microbiome abundances and multiple factors based on microbiome compositional data instead of the count data \citep{jiang2021, maity2020bayesian, altaweel2019collusivehijack}.

Recent studies have investigated the relationship between diseases and the human microbiome over time. For example, \cite{vatanen2016variation} followed 222 infants from birth to age 3 to study their gut microbiome development and its associations with the increasing incidence of autoimmune diseases. Additionally, \cite{romero2014composition} compared the vaginal microbiota of pregnant women who delivered preterm versus those who delivered at term. Longitudinal metagenomic count data is often overdispersed \citep{pramanik2016tail,hua2019assessing}, and sparse \citep{pramanik2021effects}. Two main categories exist for handling these data sets: the first involves logarithmic or other transformations on count data, followed by usage of linear mixed models to analyze the transformed data \citep{benson2010individuality,la2014patterned,leamy2014host,wang2015analysis}. The second category involves zero-inflated Gaussian mixed models to address sparsity issues in longitudinal metagenomics data \citep{zhang2020}. The first method performs poorly under certain conditions and fails to address the sparcity issue \citep{o2010not}. On the other hand, zero-inflated Gaussian mixed models successfully address the sparsity issue and can be used to analyze both transformed and untransformed metagenomic data \citep{zhang2020,pramanik2020optimization,pramanik2021optimization,pramanik2021optimal}. The second category is generalized linear mixed models, which enable direct analysis of longitudinal metagenomic count data. Metagenomic count data can typically be analyzed similarly to RNA-Seq data, assuming a negative binomial distribution \citep{pramanik2020motivation,zhang2020,pramanik2021consensus,pramanik2023scoring}.

Here, we propose a Bayesian integrative approach of computing Bayes factor to analyze microbiome count data \citep{polansky2021motif}. Our approach jointly identifies differential abundant taxa among two groups of women (i.e., pregnant and non-pregnant). The data includes 16S rRNA gene sequence based vaginal microbiota from  which samples are collected from each subject over intervals of weeks, resulting in 143 taxa and N = 900 longitudinal samples (139 measurements from pregnant women and 761 measurements from non-pregnant women.) To do our experiment we have used \cite{romero2014composition} and \cite{jiang2023flexible} data sets. Count data with large number of zeros (i.e. zero-inflation) are encountered in different fields such as medicine \citep{bohning1999}, public health \citep{zhou2000, maity2021bayesian, maity2020bayesian}, environmental sciences \citep{agarwal2002}, agriculture \citep{hall2000}, manufacturing studies\citep{lambert1992}, Orange-crowned Warblers in ponderosa \citep{garay2011,maity2022,white1996}. Zero-inflation, a common exemplification of overdispersion, refers to the incidence of zero counts is relatively higher than usual \citep{garay2011}. Since, zero counts frequently have special status in statistical literature, this definitely leads us do research in this area. For example, a production engineer might count the number of defective items selected at random from a production process \citep{bayarri2008}. If overdispersion in raw data is caused by zero-inflation, then zero-inflated Poisson (ZIP) model provides a standard framework for fitting the data \citep{garay2011,lambert1992, maity2023highest}. According to \cite{ghosh2002} when some production processes are in absolute states, zero defects occure more frequently \citep{bayarri2008}. An approach to address this issue is to use a two-parameter distribution so that the extra parameter permits a larger variance \citep{bhattacharya2008}. Double exponential family approach, a two-parametric modification of a standard one-parameter exponential family, has been developed which allows more variability than permitted by the single-parameter version \citep{bhattacharya2008,efron1986}. This is reasonable in count data distributions, such as Poisson, but not useful to model data inflated with zeros \citep{bhattacharya2008}. Fundamental idea of ZIP model is to mix a distribution degenerate at zero with a Poisson distribution \citep{garay2011}. In other words, ZIP assumes that a population consists of two individual types whereas the first type gives a zero count and the second type gives a Poisson-distributed count \citep{ridout2001, maity2019integration, maity2016bayesian}.

If a data set with zero-inflation exhibits overdispersion, a zero-inflated negative binomial (ZINB) model, mixture of a distribution degenerate at zero with a baseline negative binomial distribution, over the ZIP model \citep{garay2011}. Since overdispersion is a ramification of excess zeros, the result has excess variability and ZIP model might not a good fit for such data \citep{garay2011}. A multivariate random-parameter ZINB regression model for modeling crash counts has been developed in \cite{dong2014}. A score test for conducting hypothesis testing of ZIP regression models versus ZINB has been performed in\cite{ridout2001, paul2018bayesian}. A ZINB framework with a Gaussian process has been introduced by \cite{li2021} to analyze spatial transcriptomics data in which analysis was conducted under a Bayesian framework \citep{nam2022, calvo2023multi}. Jiang et al. \citep{jiang2021, maity2019intservbin} have been used a ZINB regression model to perform an integrative analysis on microbiome data \citep{nam2022}. In \cite{nam2022}  a statistical inference has been discussed for a zero-inflated binomial distribution with an objective Bayesian and frequentist approaches to determine a point and an interval estimators of the model parameters. Furthermore, a hypothesis testing  for excessive zeros in a zero-inflated binomial distribution have been performed and finally, a Monte Carlo simulation is utilized to investigate the performance of estimation and hypothesis testing procedures \citep{nam2022}.

Since the baseline Poisson fails to incorporate the remaining overdispersion not accounted for through zero-inflation and negative binomial models are more flexible than their Poisson counterparts in dealing with overdispersion, ZIP model is not a good fit for such data \citep{garay2011,lawless1987, maity2018power, maity2023jeffreys, maity2021circadian, maity2021semmcmc}. Moreover, it is a well known fact that the ZIP parameter estimators can be significantly biased under overdispersion of non-zero counts in relation to Poisson distribution \citep{garay2011}. Although, there is a large interest in testing of the presence of overdispersion on a given dataset, our main concentration in this paper is on those circumstances where the data exhibits overdispersion. Furthermore, in this paper we discuss Bayesian methodologies when a negative binomial (NB), ZINB, Poisson or ZIP is fitted to the dataset. We investigate the effectiveness of our theoretical results through simulation and real data analysis based on \cite{romero2014composition, ghosh2023adaptive} and \cite{jiang2023flexible} data sets. We have introduced a new R package BFZINBZIP to facilitate model selection from Poisson, NB, ZIP, and ZINB distributions.

A popular method to determine the estimates of parameters is to maximize the likelihood or natural logarithm of the likelihood with various Bayesian approaches \citep{maity2022, sommerhalder2023first}. For example, a Poisson scale representation of NB with Gamma distribution as the mixing density has been discussed in \cite{burrell1990, roy2019multiple, beck2023phase, maity2022sahpm}, a polynomial expansion and a power series expansion have been considered in \cite{bradlow2002} and \cite{bhattacharya2008} respectively. However a little has been given on the Bayesian analysis regarding ZIP versus ZINB models. In particular, to the best of our knowledge, no work exists on posterior analysis under non-informative prior analysis with above two models. We further compare our data driven results within Poisson versus NB, Poisson versus ZIP, NB versus ZINB and ZIP versus ZINB models. 

Let $\mathbf Y=[Y_1,Y_2,...,Y_n]'$ be a vector of observed count data such that each of the elements are independent and identically distributed, where $'$ represents transposition of the vector. If $\mathbf Y$ follows an NB distribution then for all positive $\gamma$ and $\kappa$, the probability density function (pdf) is defined as
\[
f^{NB}(y|\kappa)=\frac{\Gamma{(y+\gamma)}}{y!\ \Gamma{(\gamma)}}\left(1+\frac{\kappa}{\gamma}\right)^{-\gamma}\left(1+\frac{\gamma}{\kappa}\right)^{-y}, \ \ y=0,1,2,...,
\]
or, if $\mathbf Y$ follows a ZINB then for all $\a\in[0,1]$ the pdf is
\[
f^{ZINB}(y|\a,\kappa)=\begin{cases}
\a+(1-\a)\left(1+\frac{\kappa}{\gamma}\right)^{-\gamma},  &\text{if $y=0$,}\\
(1-\a)\frac{\Gamma{(y+\gamma)}}{y!\ \Gamma{(\gamma)}}\left(1+\frac{\kappa}{\gamma}\right)^{-\gamma}\left(1+\frac{\gamma}{\kappa}\right)^{-y}, &\text{if $y=1,2,...$},
\end{cases}
\]
with mean $\E(\mathbf Y)=(1-\a)\kappa$ and variance $V(\mathbf Y)=(1-\a)\kappa(1-\a\kappa+\kappa/\gamma)$, where $\a$ is represented as the zero-inflation parameter. On the other hand, if $\mathbf Y$ follows a ZIP distribution then for $\theta>0$ and a zero-inflation parameter $\hat\a\in[0,1]$ the pdf is
\[
f^{ZIP}(y|\hat\a,\theta)=\hat\a \mathbbm{1}(y=0)+(1-\hat\a)f^P(y|\theta),\ \ y=0,1,2,...,
\]
where $\mathbbm{1}(.)$ be an indicator function and $f^P(y|\theta)$ is the Poisson density so that 
\[
f^P(y|\theta)=\frac{\exp(-\theta)\ \theta^y}{y!},\ \ y=0,1,2,....
\]
Many researches have been performed using ZIP distributions with and without covariates to model count data \citep{bayarri2008}. For instance, \cite{lambert1992} and \citep{ghosh2002} have been used frequentist and Bayesian approaches respectively to explore industrial data sets through a ZIP regression model \citep{bayarri2008}. A Bayesian score test has been developed in \cite{bhattacharya2008} to test the null hypothesis $\mathcal H_0:\hat\a\leq 0$ against the alternative hypothesis $\mathcal H_1:\hat \a>0$ \citep{bayarri2008}. As frequentist approach of score test has been explained in \citep{deng2000,deng2005,van1995}, $\hat\a$ is permitted to be negative in \cite{bhattacharya2008}, as long as $\hat\a+(1-\hat\a)\exp(-\theta)\geq 0$.

In a Bayesian framework we are interested in testing
\begin{align}
\mathfrak M_0^P:Y_i\overset{\mathrm{iid}}\sim f^P(y_i|\theta),\ \ & \text{versus}\ \  \mathfrak M_1^{NB}:Y_i\overset{\mathrm{iid}}\sim f^{NB}(y_i|\kappa),\ i=1,...,n,\label{0}\\
\mathfrak M_0^P:Y_i\overset{\mathrm{iid}}\sim f^P(y_i|\theta),\ \ & \text{versus}\ \  \mathfrak M_1^{ZIP}:Y_i\overset{\mathrm{iid}}\sim f^{ZIP}(y_i|\hat\a,\theta),\ i=1,...,n,\label{1}\\
\mathfrak M_0^{NB}:Y_i\overset{\mathrm{iid}}\sim f^{NB}(y_i|\kappa),\ \ & \text{versus}\ \  \mathfrak M_1^{ZINB}Y_i\overset{\mathrm{iid}}\sim f^{ZINB}(y_i|\a,\kappa), \ i=1,...,n,\label{2}\\
\mathfrak M_0^{ZIP}:Y_i\overset{\mathrm{iid}}\sim f^{ZIP}(y_i|\hat\a,\theta),\ \ & \text{versus}\ \  \mathfrak M_1^{ZINB}Y_i\overset{\mathrm{iid}}\sim f^{ZINB}(y_i|\a,\kappa), \ i=1,...,n,\label{3}
\end{align}
where $f^P, f^{NB}, f^{ZIP}$ and $f^{ZINB}$ are the densities of Poisson, NB, ZIP and ZINB distributions, respectively, and $\mathfrak M_k^{[.]}:Y_i$ has density $f^{[.]}(y_i|\Theta_k),\Theta_k=\{\a,\hat\a,\kappa,\theta\}$ with $\Theta_k$ being the parameters in model $\mathfrak M_k^{[.]}$ for all $k=0,1$ and $[.]$ represents any of Poisson, NB, ZIP and ZINB distributions based on the testing of hypotheses in \ref{0}-\ref{3}.

The article is structured as follows: we first discuss convergence properties of the posterior distribution in Section 2. Next, we determine objective priors for the four distributions previously mentioned. Finally, we compute Bayes factors for hypotheses \ref{0}-\ref{3} and evaluate model performance on simulated data in Section 3. Two real-data analyses are presented in Section 4, and our conclusions are in Section 5.

\section{Framework}  \label{section_model}
The Bayesian methodology for choosing between $\mathfrak M_0^{[.]}$ and $\mathfrak M_1^{[.]}$ is determined by assessing the prior probabilities of each
model, the prior distributions for the model parameters, and then by computing the posterior probabilities of each $\mathfrak M_k^{[.]}$ for all $k=0,1$ \citep{bayarri2008}. The posterior probabilities are calculated from the prior distributions and the Bayes Factor, a ratio of maximum likelihood for $\mathfrak M_0^{[.]}$ and $\mathfrak M_1^{[.]}$ which is standard method in Bayesian testing and model selection and is associated with Schwarz Bayesian information criterion (BIC) \citep{bayarri2008,maity2022}.  Most of the times, due to scarcity of resources or lack of time, it is impossible to assess all the priors diligently in a subjective manner \citep{berger2006}. In this environment, objective Bayesian approach based upon non-external information (other than constructing the problem) gives a competent answer \citep{bayarri2008,berger2006}.

\subsection{An overview of the Bayes factor in model selection}

Let there be $K+1$ models $\mathfrak M_0^{[.]},....,\mathfrak M_K^{[.]}$ so that $k=0,1,...,K$ and $K>0$, and these models contend with each other in determining the most relevant model. If model $\mathfrak M_k^{[.]}$ holds, then for a parametric space $\bm \Theta_k$ of $\Theta_k$ such that $\Theta_k\subseteq\bm\Theta_k$, $\mathcal P_{\Theta_k}$ is a probability measure on a measurable space $(\mathbf Y,\mathcal A)$ such that for each $A\in\mathcal A$, $\Theta_k\mapsto\mathcal P_{\Theta_k}(A)$ is Borel measurable \citep{ghosh2010}, $Y_i$ follows a parametric distribution with pdf $f^{[.]}(y_i|\Theta_k),\Theta_k=\{\a,\hat\a,\gamma,\kappa,\theta\}$. It is convenient that $Y_1,Y_2,...$ as the coordinate random variable defined on the sample space $\Omega=\left(\mathbf Y^\oo,\mathcal A^\oo\right)$ and $\mathcal P_{\Theta_k}^\oo$ as the i.i.d. product measure defined on $\Omega$ \citep{ghosh2010}. Define the space $\Omega_n:=\left(\mathbf Y^n,\mathcal A^n\right)$ and $\mathcal P_{\Theta_k}^n$ be the $n$-fold product of $\mathcal P_{\Theta_k}$. Bayesian
model selection proceeds by choosing a prior density $\pi\left(\Theta_k|\mathfrak M_k^{[.]}\right)$ under model $\mathfrak M_k^{[.]}$ for a set of parameters $\Theta_k$, and the prior model probability $\tilde\pi\left(\mathfrak M_k^{[.]}\right)$ of $\mathfrak M_k^{[.]}$ before data $\mathbf y$ are observed so that $\mathbf y=[y_1,...,y_n]'$. Therefore, the marginal or predictive likelihood corresponding to model $\mathfrak M_k^{[.]}$ is defined as
\[
m\left(\mathbf y\big|\mathfrak M_k^{[.]}\right):=\int_{\Theta_k} f^{[.]}\left(\mathbf y\big|\Theta_k,\mathfrak M_k^{[.]}\right)\pi\left(\Theta_k|\mathfrak M_k^{[.]}\right)d\Theta_k,
\]
where $f^{[.]}\left(\mathbf y\big|\Theta_k,\mathfrak M_k^{[.]}\right)$ is the likelihood function under model $\mathfrak M_k^{[.]}$. Therefore, the posterior probability under the assumption that model $\mathfrak M_k^{[.]}$ is true can written by the following expression
\[
p\left(\mathfrak M_k^{[.]}\big|\mathbf y\right)=\frac{m\left(\mathbf y\big|\mathfrak M_k^{[.]}\right)\tilde\pi\left(\mathfrak M_k^{[.]}\right)}{\sum_{\bar k=0}^Km\left(\mathbf y\big|\mathfrak M_{\bar k}^{[.]}\right)\tilde\pi\left(\mathfrak M_{\bar k}^{[.]}\right)}\propto m\left(\mathbf y\big|\mathfrak M_k^{[.]}\right)\tilde\pi\left(\mathfrak M_k^{[.]}\right).
\]

\begin{definition}
For all n, let $p\left(\mathfrak M_k^{[.]}\big|\mathbf y\right)$ be a posterior probability for given values $y_1,y_2,...,y_n$. The sequence\\ $\left\{p\left(\mathfrak M_k^{[.]}\big|\mathbf y\right)\right\}_{i=1}^n$ is said to be consistent at $\bar\Theta_k$ if there exists a $\Omega^*\subset\Omega$ with $\mathcal P_{\bar\Theta_k}^\oo=1$ so that if $\omega\in\Omega^*$, then for every neighborhood $\mathcal N$ of $\bar\Theta_k$,
\[
p_{\mathcal N}\left(\mathfrak M_k^{[.]}\big|\mathbf y(\omega)\right)\ra 1.
\]
\end{definition}

\begin{remark}
When the metric space $\bm\Theta_k:=\left\{\Theta_k:\tilde\rho\left(\Theta_k,\bar\Theta_k\right)<1/n:n\geq1\right\}$ constructs a base for the neighborhood of $\bar\Theta_k$, and therefore it can be allowed to bet set of measure 1 to depend upon $\mathcal N$ \citep{ghosh2010}. Hence, it is enough to show that for each $\mathcal N$ of $\bar\Theta_k$, $p_{\mathcal N}\left(\mathfrak M_k^{[.]}\big|\mathbf y(\omega)\right)\ra 1$ almost everywhere of $\mathcal P_{\bar\Theta_k}^\oo$.
\end{remark}

\begin{lem}\label{l0}
Let $\left\{\mathcal P_{\Theta_k}:\Theta_k\in\bm\Theta_k\right\}$ be a probability measure dominated by a $\sigma$-finite measure $\mu$ and $f^{[.]}(y_i\big|\Theta_k)$ be the density of $\mathcal P_{\Theta_k}$. Assume $\bar\Theta_k$ be an interior point of $\bm\Theta_k$ (i.e. $\bar\Theta_k= \bm\Theta_k^o$)	and $\pi\left(\Theta_0\big|\mathfrak M_0^{[.]}\right)$, $\pi\left(\Theta_1\big|\mathfrak M_1^{[.]}\right)$ be two continuous prior densities w.r.t. a measure $\zeta$. If posterior densities $p\left(\mathfrak M_k^{[.]}\big|\mathbf y\right)$, $k=0,1$ are both consistent at $\bar\Theta_k$ then
\[
\lim_{n\ra\oo}\int_{\bm\Theta_k}\left|p\left(\mathfrak M_0^{[.]}\big|\mathbf y\right)-p\left(\mathfrak M_1^{[.]}\big|\mathbf y\right)\right|d\zeta(\Theta_k)=0,\ \text{almost sure}\ \mathcal P_{\bar\Theta_k}.
\]
\end{lem}

\begin{proof}
	See the Appendix.
\end{proof}

\begin{lem}\label{l00}
Let $p\left(\mathfrak M_0^{[.]}\big|\mathbf y\right)$ and $p\left(\mathfrak M_1^{[.]}\big|\mathbf y\right)$ be posterior probabilities for give values of $y_1,y_2,...,y_n$ such that for some constants $\Psi_k$, $\varphi_k$, $k=0,1,$
\[
1-p\left(\mathfrak M_k^{[.]}\big|\mathbf y\right)=\Psi_k\ \mathbf y^{\varphi_k}\exp(-\mathbf y^2/2)(1+o(1)),\ \text{as $\mathbf y\ra\infty$,}
\]
Then their convolution can be written as
\[
1-p\left(\mathfrak M_0^{[.]}\big|\mathbf y\right)*p\left(\mathfrak M_1^{[.]}\big|\mathbf y\right)=\sqrt{\pi}\Psi_0\Psi_1 2^{-(\varphi_1+\varphi_2)}\mathbf y^{\varphi_1+\varphi_2+1}\exp(-\mathbf y^2/4)(1+o(1)),
\] 
\end{lem}
as $\mathbf y\ra\infty$.

\begin{proof}
	See the Appendix.
\end{proof}

Along the way of \cite{ghosh2010} we will discuss \emph{Bernstein-von Mises} theorem on the asymptotic normality of the posterior distribution $p\left(\mathfrak M_k^{[.]}\big|\mathbf y\right)$. If a consistent global likelihood estimator $\Theta_k^n$ exists, then under differentiability it is easy to verify that for all $\mathcal P_{\Theta_k}$, it is a consistent solution of the likelihood equation a.s. $\mathcal P_{\Theta_k}$ \citep{ghosh2010}. To show the consistency of the posterior distribution we need the following assumption \ref{a0} of the density function $f^{[.]}\left(y_i\big|\Theta_k\right)$.

\begin{as}\label{a0}
(i). For model $\mathfrak M_k^{[.]}$, $\left\{y_i:f^{[.]}\left(y_i\big|\Theta_k\right)>0,\ i=1,...,n\right\}$ takes the same value for all $\Theta_k\in\bm\Theta_k$.\\
(ii). Suppose the likelihood function under model $\mathfrak M_k^{[.]}$ is defined as $f^{[.]}\left(\mathbf y\big|\Theta_k,\mathfrak M_k^{[.]}\right):=\ln f^{[.]}\left(\mathbf y\big|\Theta_k\right)$ for all $\mathbf y=\{y_1,...,y_n\}$ is thrice differentiable with respect to $\Theta_k$ in the neighborhood of $(\bar\Theta_k-\delta,\bar\Theta_k+\delta)$	so that 
\begin{align*}
f_1^{[.]}\left(\mathbf y\big|\Theta_k,\mathfrak M_k^{[.]}\right):=\frac{\partial}{\partial \Theta_k}f^{[.]}\left(\mathbf y\big|\Theta_k,\mathfrak M_k^{[.]}\right);\ & f_2^{[.]}\left(\mathbf y\big|\Theta_k,\mathfrak M_k^{[.]}\right):=\frac{\partial^2}{\partial \Theta_k^2}f^{[.]}\left(\mathbf y\big|\Theta_k,\mathfrak M_k^{[.]}\right);\\ \text{and,}\ f_3^{[.]}\left(\mathbf y\big|\Theta_k,\mathfrak M_k^{[.]}\right):=&\frac{\partial^3}{\partial \Theta_k^3}f^{[.]}\left(\mathbf y\big|\Theta_k,\mathfrak M_k^{[.]}\right).
\end{align*}
Then the expectations at $\bar\Theta_k$ corresponding to likelihood are $\E_{\bar\Theta_k}\left[f_1^{[.]}\left(\mathbf y\big|\Theta_k,\mathfrak M_k^{[.]}\right)\right]<\oo$ and $\E_{\bar\Theta_k}\left[f_2^{[.]}\left(\mathbf y\big|\Theta_k,\mathfrak M_k^{[.]}\right)\right]<\oo$ with
\[
\sup_{\Theta_k\in(\bar\Theta_k-\delta,\bar\Theta_k+\delta)}\left|f_3^{[.]}\left(\mathbf y\big|\Theta_k,\mathfrak M_k^{[.]}\right)\right|<L(\mathbf y);\ \text{and,}\ \E_{\bar\Theta_k}(L)<\oo.
\]
(iii). After interchange the order of expectation w.r.t. $\bar\Theta_k$ and differentiating $f^{[.]}\left(\mathbf y\big|\Theta_k,\mathfrak M_k^{[.]}\right)$ w.r.t. $\bar\Theta_k$ such that, $\E_{\bar\Theta_k}\left[f_1^{[.]}\left(\mathbf y\big|\Theta_k,\mathfrak M_k^{[.]}\right)\right]=0$ and $ \E_{\bar\Theta_k}\left[f_2^{[.]}\left(\mathbf y\big|\Theta_k,\mathfrak M_k^{[.]}\right)\right]=-\E_{\bar\Theta_k}\left[f_1^{[.]}\left(\mathbf y\big|\Theta_k,\mathfrak M_k^{[.]}\right)\right]^2$.\\
(iv). Fisher information set $\mathcal I\left(\bar\Theta_k\right):=\E_{\bar\Theta_k}\left[f_1^{[.]}\left(\mathbf y\big|\Theta_k,\mathfrak M_k^{[.]}\right)\right]^2>0$.\\
(v). Define $f^{n[.]}\left(\mathbf y\big|\Theta_k,\mathfrak M_k^{[.]}\right):=\sum_{i=1}^nf^{[.]}\left(y_i\big|\Theta_k,\mathfrak M_k^{[.]}\right)$. Then for all $\delta>0$, there $\exists\ \epsilon>0$ so that
\[
\mathcal P_{\bar\Theta_k}\left\{\sup_{\left|\Theta-\bar\Theta_k\right|>\delta}n^{-1}\left[f^{n[.]}\left(\mathbf y\big|\Theta_k,\mathfrak M_k^{[.]}\right)-f^{n[.]}\left(\mathbf y\big|\bar\Theta_k,\mathfrak M_k^{[.]}\right)\right]\leq-\epsilon\right\}\ra 1.
\]
(vi). The prior density $\pi\left(\Theta_k\big|\mathfrak M_k^{[.]}\right)$ under model $\mathfrak M_k^{[.]}$ is Lebesgue measurable, continuous and positive at $\bar\Theta_k$.
\end{as}

\begin{prop}\label{p0}
For model $\mathfrak M_k^{[.]}$ consider the density $\left\{f^{[.]}\left(\mathbf y\big|\Theta_k\right);\ \Theta_k\in\bm\Theta_k\right\}$ for all $k=1,...,K$ satisfies Assumption \ref{a0}. Let $p\left(\tau,\mathfrak M_k^{[.]}\big|\mathbf y\right)$ be the posterior density of $\tau=\sqrt{n}\left[\Theta_k-\Theta_k^n(\mathbf y)\right]$ under model $\mathfrak M_k^{[.]}$. Then
\[
\int_{\mathbb R}\left|p\left(\tau,\mathfrak M_k^{[.]}\big|\mathbf y\right)-\sqrt{\frac{\mathcal I\left(\bar\Theta_k\right)}{2\pi}}\exp\left\{-\frac{1}{2}\tau^2\mathcal I\left(\bar\Theta_k\right)\right\}\right|d\tau \overset{\mathcal P_{\bar\Theta_k}}{\to}0.
\]
\end{prop}

\begin{proof}
	See the Appendix.
\end{proof}

For a given data set $\mathbf y$ the model with the largest posterior probability is the most favorable model \citep{nam2022}. Moreover, the Bayes factor for model $\mathfrak M_k^{[.]}$ with respect to $\mathfrak M_l^{[.]}$ can be expressed as
\[
\be_{kl}(\mathbf y)\triangleq\frac{m\left(\mathbf y\big|\mathfrak M_k^{[.]}\right)}{m\left(\mathbf y\big|\mathfrak M_l^{[.]}\right)}=\frac{\int_{\Theta_k} f^{[.]}\left(\mathbf y\big|\Theta_k,\mathfrak M_k^{[.]}\right)\pi\left(\Theta_k|\mathfrak M_k^{[.]}\right)d\Theta_k}{\int_{\Theta_l} f^{[.]}\left(\mathbf y\big|\Theta_l,\mathfrak M_l^{[.]}\right)\pi\left(\Theta_l|\mathfrak M_l^{[.]}\right)d\Theta_l}.
\]
Although we have four models corresponding to Poisson, NB, ZIP and ZINP distributions, we are going to test two models at a time as written in hypotheses \ref{0}-\ref{3}. Therefore, throughout this paper $K=1$ (hence, $K+1=2$). The Bayes factor of model $\mathfrak M_1^{[.]}$ with respect to $\mathfrak M_0^{[.]}$ is 
\[
\be_{10}(\mathbf y)\triangleq\frac{m\left(\mathbf y\big|\mathfrak M_1^{[.]}\right)}{m\left(\mathbf y\big|\mathfrak M_0^{[.]}\right)}=\frac{\int_{\Theta_1} f^{[.]}\left(\mathbf y\big|\Theta_1,\mathfrak M_1^{[.]}\right)\pi\left(\Theta_1|\mathfrak M_1^{[.]}\right)d\Theta_1}{\int_{\Theta_0} f^{[.]}\left(\mathbf y\big|\Theta_0,\mathfrak M_0^{[.]}\right)\pi\left(\Theta_0|\mathfrak M_0^{[.]}\right)d\Theta_0}.
\]
Since each hypothesis consists of two models, we have $\tilde\pi\left(\mathfrak M_1^{[.]}\right)=1-\tilde\pi\left(\mathfrak M_0^{[.]}\right)$ and
\[
p\left(\mathfrak M_0^{[.]}\big|\mathbf y\right)=\frac{1}{1+\be_{10}(\mathbf y)\frac{\tilde\pi\left(\mathfrak M_1^{[.]}\right)}{\tilde\pi\left(\mathfrak M_0^{[.]}\right)}}.
\]
Furthermore, we choose model $\mathfrak M_0^{[.]}$ as the true model if 
\[
p\left(\mathfrak M_0^{[.]}\big|\mathbf y\right)=\frac{1}{1+\be_{10}(\mathbf y)\frac{\tilde\pi\left(\mathfrak M_1^{[.]}\right)}{\tilde\pi\left(\mathfrak M_0^{[.]}\right)}}>\frac{1}{2}\implies \be_{10}(\mathbf y)\frac{\tilde\pi\left(\mathfrak M_1^{[.]}\right)}{\tilde\pi\left(\mathfrak M_0^{[.]}\right)}<1\implies \be_{10}(\mathbf y)<1,
\]
and choose model $\mathfrak M_{1}^{[.]}$ as true model if $\be_{10}(\mathbf y)>1$. Following \cite{kass1992} and \cite{wasserman2000} using non-informative prior (will be discussed in the next section) yields a general interpretation of Bayes factors as given in Table \ref{table_Bayes_factor}.

\begin{table}[H]
	\centering
	\caption{Bayes Factors interpretation based upon Jeffreys' Prior. \label{table_Bayes_factor}} 
	\begin{tabular}{ |p{3cm}p{5.5cm}| }
		\hline
		\multicolumn{2}{|c|}{Bayes Factors with their meanings.} \\
		\hline
		Bayes Factor& Description\\
		\hline
 $\be_{10}(\mathbf y)<\frac{1}{10}$ & Strong presence of model $\mathfrak M_0^{[.]}$.\\
 $\frac{1}{10}\leq\be_{10}(\mathbf y)<\frac{1}{3.2}$ & Moderate presence of model $\mathfrak M_0^{[.]}$.\\
 $\frac{1}{3.2}\leq\be_{10}(\mathbf y)<1$ & Weak presence of model $\mathfrak M_0^{[.]}$.\\
$1\leq\be_{10}(\mathbf y) <3.2$ & Weak presence of model $\mathfrak M_1^{[.]}$.\\
$3.2\leq\be_{10}(\mathbf y) <10$ & Moderate presence of model $\mathfrak M_1^{[.]}$.\\
$\be_{10}(\mathbf y) >10$ & Strong presence of model $\mathfrak M_1^{[.]}$.\\
 \hline
\end{tabular}
\end{table}

\subsection{Objective priors in models with Poisson, NB, ZIP and ZINB distributions}

A severe problem in Bayesian analysis is to choose an appropriate prior $\pi\left(\Theta_k\big|\mathfrak M_k^{[.]}\right)$ under model $\mathfrak M_k^{[.]}$. The subjective Bayesian inference theory suggests that $\pi\left(\Theta_k\big|\mathfrak M_k^{[.]}\right)$ should be based on a person's prior opinion on $\Theta_k$ \citep{wasserman2000}. More common Bayesian model selection approach is based on objective theory where $\pi\left(\Theta_k\big|\mathfrak M_k^{[.]}\right)$ is chosen to be \emph{noninformative} in some sense \citep{wasserman2000}. A philosophical thinking behind this approach can be found in \cite{kass1996}. It is well known that if the common parameters are \emph{orthogonal} to the rest of the parameters in each of the $K$ models, they can be assigned the same prior distribution since the \emph{Fisher Information matrix} is block diagonal.\citep{bayarri2008,jeffreys1961,kass1992}. Since the arbitrary constant would be canceled in the Bayes factor, we use \emph{noninformative} (or \emph{improper}) prior in our case. A widely recognized noninformative prior is \emph{Jeffreys' prior}, defined as $\pi\left(\Theta_k\big|\mathfrak M_k^{[.]}\right)\propto\left| I\left(\bar\Theta_k\right)\right|^{1/2}$. In this case $\mathcal I\left(\bar\Theta_k\right):=\E_{\bar\Theta_k}\left[f_1^{[.]}\left(\mathbf y\big|\Theta_k,\mathfrak M_k^{[.]}\right)\right]^2>0$ is the Fisher information matrix as defined in Assumption \ref{a0}. For example, if $\mathbf Y\sim N(\Theta_k,I)$ then Jeffreys' prior is a flat prior $\pi\left(\Theta_k\big|\mathfrak M_k^{[.]}\right)\sim 1$, where $I$ and $N(,)$ represent an identity matrix and a multivariate normal distribution, respectively \citep{wasserman2000}.

Since $\hat\a$ and $\theta$ in ZIP are not orthogonal, following \cite{bayarri2008} with $\hat\a^*=\hat\a+(1-\hat\a)\exp(-\theta)$ the density function $f^{ZIP}(y|\hat\a,\theta)$ can be reparametrized as
\begin{equation}\label{10}
f_*^{ZIP}(y|\hat\a^*,\theta)=\hat\a^* \mathbbm{1}(y=0)+(1-\hat\a^*)f_T^P(y|\theta),\ \ y=0,1,2,...,
\end{equation}
where $f_T^P(y|\theta):=P(Y=y|Y>0;\theta)=\theta^y/\{y![\exp(\theta)-1]\}$ is the zero-truncated Poisson distribution with parameter $\theta$ such that $\hat\a^*\geq \exp(-\theta)$. Therefore, the expression for model $\mathfrak M_0^P$ is 
\begin{equation}\label{11}
f_*^{P}(y|\theta)=\exp(-\theta)\mathbbm{1}(y=0)+[1-\exp(-\theta)]f_T^P(y|\theta),\ \ y=0,1,2,....
\end{equation}
According to the suggestions in \cite{maity2022} with  $\a^*=\a+(1-\a)(1+\kappa/\gamma)^{-\gamma}$ and for all $\a^*\geq(1+\kappa/\gamma)^{-\gamma}$ the density function $f^{ZINB}(y|\a,\gamma,\kappa)$ can be represented as 
\begin{equation}\label{12}
f_*^{ZINB}(y|\a,\kappa)=\a^* \mathbbm{1}(y=0)+(1-\a^*) f_T^{NB}(y|\kappa),\ \ y=0,1,2,...,
\end{equation}
where 
\[
f_T^{NB}(y|\kappa):=P(Y=y|Y>0;\kappa)=\frac{\frac{\Gamma(y+\gamma)}{y!\Gamma(\gamma)}\left(1+\frac{\kappa}{\gamma}\right)^{-\gamma}\left(1+\frac{\gamma}{\kappa}\right)^{-y}}{1-\left(1+\frac{\kappa}{\gamma}\right)^{-\gamma}},
\]
is the zero-truncated negative binomial distributions with parameter $\kappa$ such that the expression for model $\mathfrak M_0^{NB}$ is 
\begin{equation}\label{13}
f_*^{NB}(y|\kappa)=\left(1+\frac{\kappa}{\gamma}\right)^{-\gamma}\mathbbm{1}(y=0)+\left[1-\left(1+\frac{\kappa}{\gamma}\right)^{-\gamma}\right]f_T^{NB}(y|\kappa),\ \ y=0,1,2,....
\end{equation}
As suggested by \cite{bayarri2008}, Jeffreys prior can be used for the common parameter and a \emph{proper} prior for the extra parameters. It is well known that the Jeffreys prior for $\theta$ in Poisson model, and for $\gamma$ and $\kappa$ in negative binomial model are $\pi_J^P\left(\theta\big|\mathfrak M_k^{P}\right)=1/\sqrt{\theta}$ and $\pi_J^{NB}\left(\kappa\big|\mathfrak M_k^{NB}\right)=\sqrt{\gamma/[\kappa(\gamma+\kappa)]}$, respectively \citep{bayarri2008,maity2022}. The Jeffreys prior for orthogonal ZIP (i.e., the Jeffreys prior of $f_T^P(y|\theta)$) can be expressed as
\[
\pi_J^{ZIP}\left(\theta\big|\mathfrak M_k^{ZIP}\right)=\frac{c_1(\theta)}{\sqrt{\theta}},\ \text{where}\ \ c_1(\theta)=\frac{\sqrt{1-(1+\theta)\exp(-\theta)}}{1-\exp(-\theta)}.
\]
In a similar fashion we can determine the Jeffreys prior for orthogonal ZINB (i.e., the Jeffreys prior of $f_T^{NB}(y|\kappa)$) can be expressed as
\[
\pi_J^{ZINB}\left(\kappa\big|\mathfrak M_k^{ZINB}\right)=c_2(\kappa)\sqrt{\frac{\gamma}{\kappa(\kappa+\gamma)}},
\]
where,
\begin{equation} \label{14}
c_2(\kappa)=\sqrt{\frac{\gamma^2}{\kappa^2\left[1-\left(1+\frac{\kappa}{\gamma}\right)^{-\gamma}\right]}\left[2-\frac{(\kappa+\gamma)^{-1}}{1-\left(1+\frac{\kappa}{\gamma}\right)^{-\gamma}}\right]-\frac{\kappa}{\kappa+\gamma}+\frac{\kappa(\kappa+\gamma)\left(1+\frac{\kappa}{\gamma}\right)^{-(2+\gamma)}}{\gamma\left[1-\left(1+\frac{\kappa}{\gamma}\right)^{-\gamma}\right]}\left[1+\frac{1}{\gamma}+\frac{1}{1-\left(1+\frac{\kappa}{\gamma}\right)^{-\gamma}}\right]},
\end{equation}
for all $\kappa,\gamma>0$. The derivation of Equation (\ref{14}) is presented in the Appendix. Since we need to choose a single prior for both of the NB and ZINB, and as \cite{maity2022} yields that working with any of $\pi_J^{NB}\left(\kappa\big|\mathfrak M_k^{NB}\right)$ and $\pi_J^{ZINB}\left(\a\big|\kappa,\mathfrak M_k^{ZINB}\right)$ will add negligible error in computing, we are going to choose the simpler prior version of $\pi_J^{NB}\left(\kappa\big|\mathfrak M_k^{NB}\right)=\sqrt{\gamma/[\kappa(\gamma+\kappa)]}$ for both of the NB and ZINB cases. In a similar fashion the simpler prior $\pi_J^P\left(\theta\big |\mathfrak M_k^{P}\right)=1/\sqrt{\theta}$ can be used for Poisson and ZIP cases \citep{bayarri2008}. Under orthogonal ZIP model a proper prior for $\hat\a^*$ given $\theta$ is a uniform distribution over $(\exp(-\theta),1)$ is 
\[
\pi_J^{ZIP}\left(\hat\a^*\big|\theta,\mathfrak M_k^{ZIP}\right)=\frac{1}{\sqrt{\theta}}\mathbbm{1}\left[\exp(-\theta)<\hat\a^*<1\right], \ \text{and furthermore,}\ 
\pi_J^{ZIP}\left(\hat\a\big|\theta,\mathfrak M_k^{ZIP}\right)=\frac{1}{\sqrt{\theta}}\mathbbm{1}\left[0<\hat\a<1\right].
\]
Similarly, for an orthogonal ZINB model, a proper prior for $\a^*$ given $\kappa$ is a uniform distribution over the interval $((1+\kappa/\gamma)^{-\gamma},1)$ it can be expressed as
\[
\pi_J^{ZINB}\left(\a^*\big|\kappa,\mathfrak M_k^{ZINB}\right)=\sqrt{\frac{\gamma}{\kappa(\gamma+\kappa)}}\mathbbm{1}\left[\left(1+\frac{\kappa}{\gamma}\right)^{-\gamma}<\a^*<1\right]\ \text{or,}\ \pi_J^{ZINB}\left(\a\big|\kappa,\mathfrak M_k^{ZINB}\right)=\sqrt{\frac{\gamma}{\kappa(\gamma+\kappa)}}\mathbbm{1}\left[0<\a<1\right].
\] 

\medskip

\subsection{Objective Bayes factor in models with Poisson, NB, ZIP and ZINB distributions}

\medskip

In this section we are going to determine objective Bayes factors for each of the models explained in \ref{0}-\ref{3}. For a sample of $n$ counts let $\varpi=\sum_{i=1}^n\mathbbm{1}(Y_i=0)$ be the number of zero observations, and $\varphi=\sum_{i=1}^n Y_i$ be the total count. It is important to note that $(\varpi=n)\equiv(\varphi=0)$ \citep{bayarri2008}. Therefore, by \cite{bayarri2008} for given data set $\mathbf y$
\[
f^P(\mathbf y|\theta)=\frac{\theta^\varphi\exp(-n\theta)}{\prod_{i=1}^ny_i!},\ \text{and}\ f^{ZIP}(\mathbf y|\hat\a,\theta)=\frac{[\hat\a+(1-\hat\a)\exp(-\theta)]^\varpi(1-\hat\a)^{n-\varpi}\exp\{-(n-\varpi)\theta\}\theta^\varphi}{\prod_{i=1}^ny_i!},
\]
 and by \cite{maity2022}
 \begin{align*}
 f^{NB}(\mathbf y|\kappa)&=\left(\frac{\gamma}{\gamma+\kappa}\right)^{n\gamma}\left(\frac{\kappa}{\gamma+\kappa}\right)^\varphi\left[\prod_{i=1}^n\frac{\Gamma(y_i+\gamma)}{y_i!\Gamma(\gamma)}\right],\\
 f^{ZINB}(\mathbf y|\a,\kappa)&=\left[\a+(1-\a)\left(1+\frac{\kappa}{\gamma}\right)^{-\gamma}\right]^\varpi\left(\frac{\gamma}{\gamma+\kappa}\right)^{(n-\varpi)\gamma}\left(\frac{\kappa}{\gamma+\kappa}\right)^\varphi\left[\prod_{i=1}^n\frac{\Gamma(y_i+\gamma)}{y_i!\Gamma(\gamma)}\right].
 \end{align*}
For $\varphi>0$ the marginal likelihood function of Poisson and ZIP distributions with Jeffreys priors $\pi_J^P\left(\theta\big|\mathfrak M_k^{P}\right)$ and $\pi_J^{ZIP}\left(\hat\a\big|\theta,\mathfrak M_k^{ZIP}\right)$ respectively are
\[
m\left(\mathbf y\big|\mathfrak M_k^P\right)=\frac{\Gamma\left(\varphi+\frac{1}{2}\right)}{n^{\varphi+\frac{1}{2}}\prod_{i=1}^ny_i!},\ \text{and}\ \ m\left(\mathbf y\big|\mathfrak M_k^{ZIP}\right)=\frac{\varpi!}{(n+1)!\prod_{i=1}^ny_i!}\sum_{j=0}^\varpi\frac{(n-j)!}{(\varpi-j)!}\Gamma\left(\varphi+\frac{1}{2}\right)(n-j)^{-\left(\varphi+\frac{1}{2}\right)}.
\]
On the other hand, following \citep{maity2022} the marginal likelihood function of NB distribution with Jeffreys prior $\pi_J^{NB}\left(\kappa\big|\mathfrak M_k^{NB}\right)$ is
\[
m\left(\mathbf y\big|\mathfrak M_k^{NB}\right)=\frac{1}{\sqrt{\gamma}}\left[\prod_{i=1}^n\frac{\Gamma(y_i+\gamma)}{y_i!\Gamma(\gamma)}\right]^{n\gamma+\frac{1}{2}}\text{Beta}\left(\varphi+\frac{1}{2},n\gamma\right),
\]
where ``Beta" represents a beta integration. Similarly by \cite{maity2022}, the marginal likelihood function of ZINB distribution with Jeffreys prior $\pi_J^{ZINB}\left(\a\big|\kappa,\mathfrak M_k^{ZINB}\right)$ is
\[
m\left(\mathbf y\big|\mathfrak M_k^{ZINB}\right)=\frac{\varpi !}{\sqrt{\gamma}\ (n+1)!}\left[\prod_{i=1}^n\frac{\Gamma(y_i+\gamma)}{y_i!\Gamma(\gamma)}\right]^{n\gamma+\frac{1}{2}}\sum_{j=0}^\varpi\frac{(n-j)!}{(\varpi-j)!} \ \text{Beta}\left(\varphi+\frac{1}{2},(n-j)\gamma\right).
\]
The Bayes factor of the NB model against the Poisson model (i.e., Hypothesis \ref{0}) is
\begin{equation}  \label{equation_BF_NB_Pois}
\be_{10}^1(\mathbf y)\triangleq\frac{m\left(\mathbf y\big|\mathfrak M_1^{NB}\right)}{m\left(\mathbf y\big|\mathfrak M_0^{P}\right)}=\frac{n^{\left(\varphi+\frac{1}{2}\right)}\Gamma(n\gamma)}{\sqrt{\gamma}\ \Gamma\left(\varphi+n\gamma+\frac{1}{2}\right)}\left[\prod_{i=1}^n\frac{\Gamma(y_i+\gamma)}{\Gamma(\gamma)}\right]^{n\gamma+\frac{1}{2}}\left[\prod_{i=1}^n\frac{1}{y_i!}\right]^{n\gamma-\frac{1}{2}}.
\end{equation}
It is important to note that, the Bayes factor $\be_{10}^1$ is increasing in total count $\varphi$ for any given $\gamma$ and $n$. When $\varphi=0$ or equivalently all counts are zero ($\mathbf y=0$), $\be_{10}^1(0)=n^{1/2}\Gamma(n\gamma)/[\sqrt{\gamma}\ \Gamma(n\gamma+1/2)]<\infty$. Following \cite{bayarri2008} the Bayes factor of the ZIP model against the Poisson model (i.e., Hypothesis \ref{1}) is
\[
\be_{10}^2(\mathbf y)\triangleq\frac{m\left(\mathbf y\big|\mathfrak M_1^{ZIP}\right)}{m\left(\mathbf y\big|\mathfrak M_0^{P}\right)}=\frac{\varpi !}{(n+1)!}\sum_{j=0}^\varpi\frac{(n-j)!}{(\varpi-j)!}\left(1-\frac{j}{n}\right)^{-\left(\varphi+\frac{1}{2}\right)}.
\]
\cite{bayarri2008} suggests that when $\varphi=0$, $m\left(\mathbf y=0\big|\mathfrak M_0^{P}\right)=\Gamma(1/2)/\sqrt{n}$ and $m\left(\mathbf y=0\big|\mathfrak M_1^{ZIP}\right)=\infty$ which implies $\be_{10}^2(0)=\infty$. In this case, for a given $n$, $\be_{10}^2(\mathbf y)$ is increasing in $\varphi$ for any fixed $\varpi$, and is increasing in $\varpi$ for any given $\varphi$ \citep{bayarri2008}. Now the Bayes factor of the ZINB model against the NB model (i.e., Hypothesis \ref{2}) is
\[
\be_{10}^3(\mathbf y)\triangleq\frac{m\left(\mathbf y\big|\mathfrak M_1^{ZINB}\right)}{m\left(\mathbf y\big|\mathfrak M_0^{NB}\right)}=\frac{\varpi!\Gamma\left(\varphi+n\gamma+\frac{1}{2}\right)}{(n+1)!\Gamma(n\gamma)}\sum_{j=0}^\varpi\frac{(n-j)!\Gamma((n-j)\gamma)}{(\varpi-j)!\Gamma\left(\varphi+(n-j)\gamma+\frac{1}{2}\right)}.
\]
For any give $n$ and $\gamma$ if $\varphi=0$ then, $\be_{10}^3(0)=\sum_{j=0}^n\Gamma((n-j)\gamma)\Gamma(n\gamma+1/2)/[n\Gamma(n\gamma)\Gamma((n-j)\gamma+1/2)]<\infty$. Finally, the Bayes factor of the ZINB model against the model ZIP (i.e., Hypothesis \ref{3}) is
\[
\be_{10}^4(\mathbf y)\triangleq\frac{m\left(\mathbf y\big|\mathfrak M_1^{ZINB}\right)}{m\left(\mathbf y\big|\mathfrak M_0^{ZIP}\right)}=\frac{1}{\sqrt{\gamma}}\left[\prod_{i=1}^n\frac{\Gamma(y_i+\gamma)}{\Gamma(\gamma)}\right]^{n\gamma+\frac{1}{2}}\left[\prod_{i=1}^n\frac{1}{y_i!}\right]^{n\gamma-\frac{1}{2}}\sum_{j=0}^\varpi\frac{\Gamma((n-j)\gamma)}{(n-j)^{-\left(\varphi+\frac{1}{2}\right)}\Gamma\left(\varphi+(n-j)\gamma+\frac{1}{2}\right)}.
\]
It can be easily verified that for any given $n$ and $\gamma$, $\be_{10}^4(\mathbf y)$ is strictly increasing in $\varphi$. Furthermore, when $\varphi=0$,  $\be_{10}^4(0)=(\gamma)^{-1/2}\sum_{j=0}^n \Gamma((n-j)\gamma)/\left[(n-j)^{-1/2}\Gamma\left((n-j)\gamma+1/2\right)\right]<\infty$.

\section{Simulation Study}

In this section we carry out a series of simulation studies to estimate some operating characteristics of the Bayes factors derived in the previous Section. 

\subsection{Bayes factor of Negative Binomial against Poisson}

In the first experiment, we generate 1000 simulated datasets from either the NB distribution or the Poisson distribution with different parameter settings. The exact values of the parameters are given in Table \ref{table_simulation_NB_Pois}. For each simulation, we compute the Bayes factor derived in Section $2.3$ that is the evidence of the ZINB distribution against the NB distribution. Note that, when computing the Bayes factor, $ \gamma $ has been assumed to be fixed as discussed in Section \ref{section_model}. Empirically, it has been noted that $ \gamma = 1.001 $ offers the best outcome. 

In the following, it is said that the Bayes factor fevers the NB model against the Poisson model if the computed log(Bayes factor) is more than log(3.2) or log(10). If the computed log(Bayes factor) is more than log(3.2) then the evidence is substantial and if the computed log(Bayes factor) is more than log(10) then it is said that there is strong evidence that the model under consideration is a NB model (see Table \ref{table_Bayes_factor}). On the other hand, if the computed log(Bayes factor) is less than log(3.2) or log(10), then the evidence is substantial or strong respectively in the favor of Poisson model. In terms of the notations introduced in Section \ref{section_model}, if we denote NB and Poisson model by $ \mathfrak {M}_1 $ and $ \mathfrak {M}_0 $ then Table \ref{table_Bayes_factor} is directly applicable to draw the inference. 

\begin{table}[!ht]
    \centering
    \caption{Simulation result to count how many times Statistical procedures favor NB model against the Poisson model. BF3: Number of times the log(Bayes factor) is more than 3.2 (when the data generating model is NB) or less than 1/3.2 (when the data generating model is Poisson), BF10: Number of times the log(Bayes factor) is more than 10 (when the data generating model is NB) or less than 1/10 (when the data generating model is Poisson), Vuong: number of times the data generating model is selected by Vuong's test, AIC: number of times the data generating model is selected by AIC criterion out of 1000 simulations.  \label{table_simulation_NB_Pois}}
    \begin{tabular}{cccccccccc}
    \hline
    \hline
        $ \lambda $ & $ \gamma $ & $ \kappa $ & Data Generating Model & BF3 & BF10 & Vuong & AIC \\ 
        \hline
        \hline
        0.5 & 1.5 & 0.5 & NB & 969 & 900 & 59 & 628 \\ 
        ~ & ~ & ~ & Pois & 66 & 17 & 123 & 928 \\ 
        ~ & 0.5 & 0.5 & NB & 1000 & 998 & 518 & 979 \\ 
        ~ & ~ & ~ & Pois & 68 & 14 & 126 & 945 \\ 
        ~ & 0.5 & 1.5 & NB & 1000 & 1000 & 999 & 1000 \\
        ~ & ~ & ~ & Pois & 75 & 19 & 126 & 941 \\ 
        \hline
        1 & 1.5 & 0.5 & NB & 972 & 897 & 46 & 597 \\ 
        ~ & ~ & ~ & Pois & 431 & 304 & 116 & 933 \\ 
        ~ & 0.5 & 0.5 & NB & 999 & 995 & 517 & 973 \\ 
        ~ & ~ & ~ & Pois & 426 & 301 & 116 & 942 \\ 
        ~ & 0.5 & 1.5 & NB & 1000 & 1000 & 995 & 1000 \\
        ~ & ~ & ~ & Pois & 393 & 263 & 100 & 935 \\ 
        \hline
        3 & 1.5 & 0.5 & NB & 961 & 893 & 63 & 608 \\ 
        ~ & ~ & ~ & Pois & 988 & 980 & 102 & 933 \\ 
        ~ & 0.5 & 0.5 & NB & 999 & 996 & 533 & 976 \\ 
        ~ & ~ & ~ & Pois & 989 & 982 & 106 & 934 \\ 
        ~ & 0.5 & 1.5 & NB & 1000 & 1000 & 996 & 1000 \\
        ~ & ~ & ~ & Pois & 992 & 986 & 58 & 942 \\ 
        \hline
        5 & 1.5 & 0.5 & NB & 963 & 929 & 63 & 619 \\ 
        ~ & ~ & ~ & Pois & 1000 & 1000 & 84 & 935 \\ 
        ~ & 0.5 & 0.5 & NB & 1000 & 998 & 519 & 978 \\ 
        ~ & ~ & ~ & Pois & 1000 & 1000 & 104 & 936 \\ 
        ~ & 0.5 & 1.5 & NB & 1000 & 1000 & 996 & 1000 \\
        ~ & ~ & ~ & Pois & 1000 & 1000 & 111 & 927 \\ 
        \hline
        \hline
    \end{tabular}
\end{table}

Table \ref{table_simulation_NB_Pois} summarizes the result how many times the zero inflated model is selected out of 1000 simulations using the Bayes factor comparisons. Additionally, we have included the outcome using the Vuong's test \citep{vuong1989likelihood} and akaiake information criterion (AIC, \citep{akaike1998information}).  \textsf{R} package \pkg{nonnest2} \citep{Merkle2020nonnest2, R} has been utilized to carry out Vuong's test. 

Nevertheless, it is evident from Table \ref{table_simulation_NB_Pois} that Bayes factor remains superior in selecting the correct model if the data generating model follows a NB distribution. It remains superior in selecting the correct model when the data generating model is a Poisson model if the mean of the Poisson distribution is high. Moreover, the criterion -- log(Bayes factor) more than log(3.2) (BF3) -- selects the zero inflated model more often than the criterion --  log(Bayes factor) more than log(10) (BF10) -- for obvious reason. For instance, with data generating $ \lambda = 5, \gamma = 1.5, \kappa = 0.5$, when the sample is simulated from a NB distribution, then BF3 and BF10 are able to recover the NB distribution 963 times and 929 times respectively. On the other hand, AIC criterion indicates that 619 datasets follows the NB model out of 1000 simulated datasets. With the same data generating parameters, when the data are simulated from a Poisson distribuion, then, BF3 and B10 are able to recover the Poisson model 1000 times and 1000 times respctively, outperfroming the AIC creterion which is able to indicate in the favor of the Poisson model 935 times. Note that, the performance of Vuong's test remains inferior throughout the simulation studies.

\subsection{Bayes factor of Zero Inflated Poisson against Poisson}

In the second experiment, we generate 1000 simulated datasets from either the zero inflated Poisson (ZIP) distribution or the Poisson distribution with different parameter settings. The exact values of the parameters are given in Table \ref{table_simulation_ZIP_Pois}. The data generation and the inference follows the similar paths as the first simulated example. 

\begin{table}[!ht]
    \centering
    \caption{Simulation result to count how many times Statistical procedures favor ZIP model against the Poisson model. BF3: Number of times the log(Bayes factor) is more than 3.2 (when the data generating model is ZIP) or less than 1/3.2 (when the data generating model is Poisson), BF10: Number of times the log(Bayes factor) is more than 10 (when the data generating model is ZIP) or less than 1/10 (when the data generating model is Poisson), Vuong: number of times the data generating model is selected by Vuong's test, Inflation: Number of times it is predicted that the data are zero inflated, AIC: number of times the data generating model is selected by AIC criterion out of 1000 simulations. Percentage (\%) of Zeros: Percentage of zeros present in the data.  \label{table_simulation_ZIP_Pois} }
    \begin{tabular}{cccccccc}
    \hline
    \hline
        $ \lambda $ & Percentage (\%) of Zeros & Data Generating Model & BF3 & BF10 & Vuong & Inflation & AIC \\ \hline
        0.5 & 97.7 & ZIP & 415 & 294 & 36 & 2 & 415 \\ 
        ~ & 60.9 & Pois & 559 & 28 & 45 & 814 & 939 \\ 
        ~ & 90.3 & ZIP & 590 & 362 & 80 & 50 & 644 \\ 
        ~ & 60.8 & Pois & 567 & 35 & 51 & 790 & 935 \\ 
        ~ & 80.3 & ZIP & 390 & 191 & 36 & 195 & 519 \\ 
        ~ & 60.6 & Pois & 578 & 23 & 38 & 789 & 929 \\ 
        ~ & 70.6 & ZIP & 135 & 48 & 9 & 178 & 258 \\ 
        ~ & 60.5 & Pois & 560 & 18 & 44 & 799 & 943 \\ 
        \hline
        1 & 96.8 & ZIP & 765 & 633 & 173 & 75 & 765 \\ 
        ~ & 37 & Pois & 810 & 349 & 46 & 388 & 922 \\ 
        ~ & 84.3 & ZIP & 937 & 859 & 508 & 743 & 958 \\ 
        ~ & 36.8 & Pois & 815 & 320 & 47 & 409 & 944 \\ 
        ~ & 68.2 & ZIP & 869 & 715 & 419 & 935 & 949 \\ 
        ~ & 36.6 & Pois & 795 & 322 & 47 & 385 & 926 \\ 
        ~ & 52.5 & ZIP & 390 & 220 & 77 & 806 & 643 \\ 
        ~ & 36.7 & Pois & 803 & 338 & 50 & 390 & 918 \\ 
        \hline
        3 & 95.2 & ZIP & 995 & 989 & 810 & 868 & 995 \\ 
        ~ & 4.9 & Pois & 959 & 853 & 72 & 207 & 926 \\ 
        ~ & 76.5 & ZIP & 1000 & 1000 & 999 & 1000 & 1000 \\
        ~ & 4.9 & Pois & 963 & 845 & 66 & 194 & 937 \\ 
        ~ & 52.5 & ZIP & 1000 & 1000 & 1000 & 1000 & 1000 \\ 
        ~ & 4.9 & Pois & 959 & 845 & 72 & 203 & 934 \\ 
        ~ & 28.6 & ZIP & 1000 & 999 & 995 & 1000 & 1000 \\ 
        ~ & 5 & Pois & 956 & 849 & 80 & 228 & 934 \\ 
        \hline
        5 & 95 & ZIP & 1000 & 1000 & 1000 & 1000 & 1000 \\ 
        ~ & 1.3 & Pois & 961 & 884 & 0 & 642 & 884 \\ 
        ~ & 74.9 & ZIP & 1000 & 1000 & 1000 & 1000 & 1000 \\ 
        ~ & 1.4 & Pois & 962 & 879 & 0 & 710 & 881 \\ 
        ~ & 50.3 & ZIP & 1000 & 1000 & 1000 & 1000 & 1000 \\ 
        ~ & 1.4 & Pois & 964 & 892 & 0 & 620 & 891 \\ 
        ~ & 25.5 & ZIP & 1000 & 1000 & 1000 & 1000 & 1000 \\ 
        ~ & 1.4 & Pois & 964 & 898 & 0 & 652 & 896 \\ 
        \hline
        \hline
    \end{tabular}
\end{table}

Table \ref{table_simulation_ZIP_Pois} summarizes the result how many times the zero inflated model is selected out of 1000 simulations using the Bayes factor comparisons, Vuong's test and the AIC criterion. An additional outcome has been included using the \textsf{R} package \pkg{performance} written by \cite{Ludecke2021performance}. This package offers functionality to check if excessive amount of zeros are present in the data. In this way, if it is determined that the number of existing zero's are than the usual then one can conclude that the data follows a zero inflated distribution. 

Nevertheless, it is evident from Table \ref{table_simulation_ZIP_Pois} that Bayes factor remains superior in selecting the correct model, particularly, when the mean of the Possion distribution is large. The other inferences remain similar to the first simulated example.

\subsection{Bayes factor of Zero Inflated Negative Binomial against Negative Binomial}

In the third experiment, we generate 1000 simulated datasets from either the zero inflated Negative Binomial (ZINB) distribution or the Negative Binomial (NB) distribution with different parameter settings. The exact values of the parameters are given in Table \ref{table_simulation_ZINB_NB}. The data generation and the inference follows the similar paths as the previous examples. 

\begin{table}[!ht]
    \centering
    \caption{Simulation result to count how many times Statistical procedures favor ZINB model against the NB model. BF3: Number of times the log(Bayes factor) is more than 3.2 (when the data generating model is ZINB) or less than 1/3.2 (when the data generating model is NB), BF10: Number of times the log(Bayes factor) is more than 10 (when the data generating model is ZINB) or less than 1/10 (when the data generating model is NB), Vuong: number of times the data generating model is selected by Vuong's test, Inflation: Number of times it is predicted that the data are zero inflated, AIC: number of times the data generating model is selected by AIC criterion out of 1000 simulations. Percentage (\%) of Zeros: Percentage of zeros present in the data.  \label{table_simulation_ZINB_NB} }
    \begin{tabular}{ccrlcccccc}
    \hline
    \hline
        $ \gamma $ & $ \kappa $ & Percentage (\%) of Zeros & Data Generating Model & BF3 & BF10 & Vuong & Inflation & AIC \\ 
        \hline
        \hline
        1.5 & 0.5 & 96.9 & ZINB & 1000 & 1000 & 54 & 32.6 & 848 \\
        ~ & ~ & 45.5 & NB & 40 & 0 & 20 & 910 & 470 \\ 
        ~ & ~ & 86.8 & ZINB & 1000 & 1000 & 60 & 0 & 820 \\ 
        ~ & ~ & 46.5 & NB & 50 & 0 & 30 & 909 & 485 \\ 
        ~ & ~ & 73.8 & ZINB & 1000 & 1000 & 120 & 0 & 870 \\ 
        ~ & ~ & 46.2 & NB & 60 & 10 & 20 & 930 & 440 \\ 
        ~ & ~ & 59.8 & ZINB & 1000 & 990 & 40 & 0 & 720 \\ 
        ~ & ~ & 47.0 & NB & 40 & 10 & 0 & 879 & 374 \\
        \hline
        0.5 & 0.5 & 97.6 & ZINB & 1000 & 1000 & 34 & 966 & 896 \\
        ~ & ~ & 57.3 & NB & 0 & 0 & 20 & 1000 & 450 \\ 
        ~ & ~ & 92.4 & ZINB & 1000 & 1000 & 20 & 0 & 808 \\ 
        ~ & ~ & 71.0 & NB & 0 & 0 & 0 & 1000 & 410 \\ 
        ~ & ~ & 85.2 & ZINB & 1000 & 1000 & 20 & 0 & 737 \\ 
        ~ & ~ & 71.0 & NB & 0 & 0 & 10 & 1000 & 420 \\ 
        ~ & ~ & 77.6 & ZINB & 1000 & 1000 & 20 & 0 & 600 \\ 
        ~ & ~ & 58.5 & NB & 0 & 0 & 10 & 1000 & 440 \\ 
        \hline
        5 & 5 & 94.9 & ZINB & 1000 & 1000 & 646 & 545 & 1000 \\ 
        ~ & ~ & 3.3 & NB & 1000 & 1000 & 20 & 265 & 510 \\ 
        ~ & ~ & 75.9 & ZINB & 1000 & 1000 & 900 & 200 & 1000 \\ 
        ~ & ~ & 3.2 & NB & 1000 & 10000 & 0 & 280 & 480 \\ 
        ~ & ~ & 51.4 & ZINB & 1000 & 1000 & 987 & 953 & 1000 \\
        ~ & ~ & 3.3 & NB & 1000 & 1000 & 41 & 301 & 499 \\ 
        ~ & ~ & 27.7 & ZINB & 910 & 860 & 930 & 1000 & 1000 \\ 
        ~ & ~ & 3.3 & NB & 1000 & 1000 & 20.8 & 271 & 521 \\ \hline
        \hline
    \end{tabular}
\end{table}

Table \ref{table_simulation_ZINB_NB} summarizes the result how many times the zero inflated model is selected out of 1000 simulations using the Bayes factor comparisons, Vuong's test, inflation, and the AIC criterion. It is evident from Table \ref{table_simulation_ZINB_NB} that Bayes factor remains superior in selecting the correct model, particularly, when the parameters of the NB distribution are large. The other inferences remain similar to the previous simulated examples.

\subsection{Bayes factor of Zero Inflated Negative Binomial against Zero Inflated Poisson}

In the last experiment, we generate 1000 simulated datasets from either the zero inflated Negative Binomial (ZINB) distribution or the zero inflated Poisson (ZIP) distribution with different parameter settings. The exact values of the parameters are given in Table \ref{table_simulation_ZINB_ZIP}. The data generation and the inference follows the similar paths as the previous examples. 

\begin{table}[!ht]
    \centering
    \caption{Simulation result to count how many times Statistical procedures favor ZINB model against the ZIP model. BF3: Number of times the log(Bayes factor) is more than 3.2 (when the data generating model is ZINB) or less than 1/3.2 (when the data generating model is NB), BF10: Number of times the log(Bayes factor) is more than 10 (when the data generating model is ZINB) or less than 1/10 (when the data generating model is NB), Vuong: number of times the data generating model is selected by Vuong's test, Inflation: Number of times it is predicted that the data are zero inflated, AIC: number of times the data generating model is selected by AIC criterion out of 1000 simulations. Percentage (\%) of Zeros: Percentage of zeros present in the data. \label{table_simulation_ZINB_ZIP} }
    \begin{tabular}{cccrlcccc}
    \hline
        $ \lambda $ & $ \gamma $ & $ \kappa $ & Percentage (\%) of Zeros & Data Generating Model & BF3 & BF10 & Vuong & AIC \\ \hline
        \hline
        1 & 0.5 & 0.5 & 84 & ZINB & 587 & 587 & 49 & 665 \\ 
        ~ & ~ & ~ & 68.1 & ZIP & 990 & 990 & 61 & 624 \\ 
        1 & 5 & 5 & 53 & ZINB & 1000 & 1000 & 278 & 994 \\ 
        ~ & ~ & ~ & 69.4 & ZIP & 989 & 989 & 45 & 603 \\ 
        \hline
        3 & 0.5 & 0.5 & 86.6 & ZINB & 579 & 579 & 41 & 665 \\ 
        ~ & ~ & ~ & 52.6 & ZIP & 518 & 518 & 101 & 556 \\ 
        3 & 5 & 5 & 53.2 & ZINB & 999 & 999 & 281 & 994 \\ 
        ~ & ~ & ~ & 54.4 & ZIP & 518 & 518 & 84 & 573 \\ 
        \hline
        \hline
    \end{tabular}
\end{table}

Table \ref{table_simulation_ZINB_ZIP} summarizes the result how many times the zero inflated model is selected out of 1000 simulations using the Bayes factor comparisons, Vuong's test, and the AIC criterion. It is evident from Table \ref{table_simulation_ZINB_ZIP} that Bayes factor remains superior in selecting the correct model.

\section{Model Selection in Microbiome Data}

In this Section we apply the Bayes factor computation techniques discussed here in a real life data originated from a case-control study. The objective of the original experiment was to gain knowledge of the vaginal microbioata throughout pregnancy. Toward this end, a longitudinal case control study was designed in 22 pregnant women who delivered at term (38 to 42 weeks) without complications, and 32 non-pregnant women. Serial samples of vaginal fluid were collected from both non-pregnant and pregnant patients. The data includes 16S rRNA gene sequence based vaginal microbiota from  which samples are collected from each subject over intervals of weeks, resulting in 143 taxa and N = 900 longitudinal samples (139 measurements from pregnant women and 761 measurements from non-pregnant women.) For more details on the experiment see \cite{romero2014composition}; also see \cite{jiang2023flexible}. 

For the analysis, we focused on two specific Phylotypes: Lactobacillus.iners and Atopobium. Each dataset contained 900 observations, with the first dataset having 15.1\% zeros and the second dataset having 66.3\% zeros. We computed the log(Bayes factor) and AIC criteria for four models: Negative Binomial (NB), Poisson, Zero-Inflated Negative Binomial (ZINB), and Zero-Inflated Poisson (ZIP).

Table \ref{table_microbiom_BF} presents the computed log(Bayes Factor) on the Microbiome data, while Table \ref{table_microbiom_aic} displays the AIC values for each model. Note that, a Negative Binomial model cannot be fitted to the data because the underlying maximization process does not converge. For the same reason, the Bayes factor of NB against Poisson model cannot be computed. 

For the first dataset, the log(Bayes factor) of ZINB against NB and of ZIP against Poisson are 829.0 and 171854.9, respectively, which favors a zero Inflated model for the data. Consequently, the log(Bayes factor) of ZINB against ZIP becomes -13686110.0 which implies that one should fit a zero inflated Poisson model to the data. On the other hand, the AIC criterion supports to fit a zero Inflated Negative Binomial model to the data. \cite{romero2014composition} concluded in the favor of fitting a negative Binomial model. 

\begin{table}[H]
    \centering
    \caption{Computed log(Bayes Factor) on the Microbiome data. \label{table_microbiom_BF}}
    \begin{tabular}{l|l|r}
    \hline
    \hline
         Example & Model & log(Bayes factor) \\
         \hline
         \hline
         1 & NB vs. Poisson  & -- \\
           & ZINB vs. NB     & 829.0 \\
           & ZIP vs. Poisson & 171854.9 \\
           & ZINB vs. ZIP    & -13686110.0 \\
           \hline
         2 & NB vs. Poisson  & -- \\
           & ZINB vs. NB     & 1172.6 \\
           & ZIP vs. Poisson & 5073.6 \\
           & ZINB vs. ZIP    & 120266.8 \\
           \hline
           \hline
    \end{tabular}
\end{table}

\begin{table}[!ht]
    \centering
    \caption{Computed log(Bayes Factor) on the Microbiome data. \label{table_microbiom_aic}}
    \begin{tabular}{l|l|r}
    \hline
    \hline
         Example & Model & AIC \\
         \hline
         \hline
         1 & NB      & -- \\
           & Poisson & 1667918.0 \\
           & ZINB    & 12513.0 \\
           & ZIP     & 1324204.0 \\
           \hline
         2 & NB      & -- \\
           & Poisson & 24913.9 \\
           & ZINB    & 3342.2 \\
           & ZIP     & 14763.3 \\
           \hline
           \hline
    \end{tabular}
\end{table}

A very similar analysis concludes that a ZINB model is the appropriate one to fit into the second dataset. This can be concluded by computing both th Bayes factor and the AIC. Furthermore, this inference accords with the findings of \cite{romero2014composition}.  

Overall, the Bayes factor and AIC analyses provide insights into selecting the appropriate models for further analysis of the vaginal microbiota data obtained from the case-control study.

\medskip

\section{Discussion}

\medskip

In recent years, a significant effort has done in the literature of longitudinal metagenomics to investigate dynamic associations between microbial symbiosis and the development of many diseases, such as inflammatory bowl diseases \citep{sharpton2017development,minar2018evaluating}, colorectal cancers \citep{liang2014dynamic}, Parkinson's disease \citep{yang2018longitudinal,minar2018grand,minar2019tatmadaw}, preterm birth \citep{stewart2017longitudinal}, and autoimmune diseases \citep{vatanen2016variation,zhang2020,roy2023prevalence,roy2023obeseye}. The literature discussed above either used 16S rRNA or whole-metagenome shotgun sequencing technologies to simulate longitudinal metagenomics data \citep{zhang2020,roy2023machine}. While the bioinformatics tools for processing 16S rRNA sequencing data give the counts, whole-metagenome shotgun sequencing data give either counts or proportions. In this article, we considered the longitudinal metagenomic count data generated from 16S rRNA sequencing based vaginal microbiota. Since the objective was to gain knowledge of the vaginal microbiota throughout pregnancy, a longitudinal case control study was designed in 22 pregnant women who delivered at term (38 to 42 weeks) without complications, and 32 non-pregnant women, and serial samples of vaginal fluid were collected from both non-pregnant and pregnant patients. Moreover, we analyzed on two specific Phylotypes: Lactobacillus.iners and Atopobium. Each dataset contained 900 observations, with the first dataset having 15.1\% zeros and the second dataset having 66.3\% zeros. We computed the log(Bayes factor) and AIC criteria for four models: Negative Binomial (NB), Poisson, Zero-Inflated Negative Binomial (ZINB), and Zero-Inflated Poisson (ZIP).

In this article, we presented Poisson, NB, ZIP, and ZINB distributions to analyze high-throughput sequencing microbiome data. First, we verified some convergence and measurability properties of the posterior distribution. Second, the Jeffreys prior was calculated for ZINB. Then the presence of over-dispersion was tested by using Bayesian methodologies. We introduced the Bayes factor for ZINB and ZIP and tested for the model selection under the incidence of over dispersed data. For each of the four distributions, we used non-informative Jeffreys prior and determined Bayes factors corresponding to the hypotheses \ref{0}-\ref{3}. We did simulation studies of the distributions with different parameters to determine the effectiveness of Bayes' factors (i.e., BF3 and BF10) compared with traditional AIC and Vuong's test. We showed that BF3 and BF10 outperformed AIC and Vuong's test in every case. For example, in the case of NB versus Poisson with $\lambda=1$, $\gamma=1.5$, and $\kappa=0.5$, when a sample is generated by simulating an NB, then BF3 and BF10 would be able to recover the NB distribution 972 and 897 times, respectively (see Table \ref{table_simulation_NB_Pois}). On the other hand, AIC indicates that 597 datasets follow the NB distribution out of 1000 simulated datasets. In this case, Vuong's test gives the most inferior result and throughout our simulation studies, its performance was the worst. To conduct the quantitative analysis, R package BFZINBZIP was used which is available at authors' github account.

Our method is novel in identifying differentially 143 taxa for two patient groups (i.e., pregnant and non-pregnant women) under a single statistical framework which allows for an integrative analysis of the microbiome and other omics data sets. The proposed method can lead to proper clinical decisions corresponding to the precision shaping of the microbiome data. Furthermore, BF3 and BF10 proposed in this article perform better than AIC and Vuong's test throughout our simulation studies and real data analysis. In real data analysis, since the underlying maximization process of 16S rRNA data do not converge, an NB distribution is impossible to fit. As a result, the Bayes factor of NB against the Poisson model cannot be determined. In Table, \ref{table_microbiom_BF}, the log(Bayes factor) of ZINB against NB, and of ZIP against Poisson are 829.0 and 171854.9, respectively, which supports the zero-inflated model for our data set. On the other hand, the log(Bayes factor) of ZINB against ZIP is -13686110.0 supports in favor of the implementation of a ZIP model to the data. Furthermore, the AIC criterion in Table \ref{table_microbiom_aic} goes in favor of fitting a ZINB to the data. Tables \ref{table_microbiom_BF}  and \ref{table_microbiom_aic} give similar results for the second set of data which favors the implementation of a ZINB model as the log(Bayes factor) and AIC are 120266.8 and 3342.2, respectively. This inference is similar to the results obtained in \cite{romero2014composition}.

The framework of the proposed method allows for several extensions. For example, the current model supports two groups (i.e., pregnant and non-pregnant women). We can extend our current model to multiple phenotype type groups including intermediate phenotypes. In this case, our method can incorporate group-specific parameters while holding other parameters fixed, and same poaterior inference can be incorporated. We can extend our proposed model to a regression framework where the normalized microbiome normalized abundance can be used as a the response which would integrate metabolite compounds, as predictors \citep{jiang2021}. Another extension would be to discuss correlated covariates such as longitudinal clinical measurements \citep{zhang2017,jiang2021}.

\section*{Conflict of Interest}
None declared.

\section*{Supplementary Material}
An R package BFZINBZIP is available on Github:\\
\url{https://github.com/arnabkrmaity/BFZINBZIP/tree/main}. This package has been used to do simulations in Section 3 and real data analysis in Section 4.

\section*{Data Availability Statement}
Romero data is available in their paper \cite{romero2014composition} or directly from the R package NBZIMM.

\section*{Appendix}

\medskip

\subsection*{Proof of Lemma \ref{l0}}
In order to prove this lemma we will show that for the probability measure at $\bar \Theta_k$ of model $\mathfrak M_{k}^{[.]}$ denoted as $\mathcal P_{\Theta_k}^\oo(\Omega^*)=1$
\[
\int_{\bm\Theta_k}p\left(\mathfrak M_1^{[.]}\big|\mathbf y(\omega)\right)\left|1-\frac{p\left(\mathfrak M_0^{[.]}\big|\mathbf y(\omega)\right)}{p\left(\mathfrak M_1^{[.]}\big|\mathbf y(\omega)\right)}\right|d\zeta(\Theta_k)\ra 0,
\]
where $p\left(\mathfrak M_k^{[.]}\big|\mathbf y(\omega)\right)$ be a posterior density distribution of $k^{th}$ model so that $k=0,1$. Using the continuity at point $\bar\Theta_k\subset\Theta_k$, there $\exists\{\delta,\epsilon,\lambda\}$ so that for all $\delta>0,\epsilon>0$ and $\lambda>0$ there exists a neighborhood $\mathcal N$ of $\bar\Theta_k$such that $\forall\ \Theta_k\in\mathcal N$, and $k=1,2$,
\[
\left|\frac{\pi\left(\Theta_0|\mathfrak M_0^{[.]}\right)}{\pi\left(\Theta_1|\mathfrak M_1^{[.]}\right)}-\frac{\pi\left(\bar\Theta_0|\mathfrak M_0^{[.]}\right)}{\pi\left(\bar\Theta_1|\mathfrak M_1^{[.]}\right)}\right|<\delta,\ \text{and}\ \left|\pi\left(\bar\Theta_k|\mathfrak M_k^{[.]}\right)-\pi\left(\Theta_k|\mathfrak M_k^{[.]}\right)\right|<\delta.
\]
By consistency there exists a sample space $\Omega$, $\mathcal P_{\bar\theta_k}^\oo(\Omega)=1$, so that for each $\omega\in\Omega$, we have the posterior probability at neighborhood $\mathcal N$ of $\Theta_k$ as
\[
p_{\mathcal N}\left(\mathfrak M_k^{[.]}\big|\mathbf y(\omega)\right)=\frac{\int_{\mathcal N}\prod_{i=1}^n f^{[.]}\left(y_i(\omega)|\bm\Theta_k,\mathfrak M_k^{[.]}\right)\pi\left(\Theta_k|\mathfrak M_k^{[.]}\right)d\zeta(\Theta_k)}{\int_{\bm\Theta_k}\prod_{i=1}^nf^{[.]}\left(y_i(\omega)|\bm\Theta_k,\mathfrak M_k^{[.]}\right)\pi\left(\Theta_k|\mathfrak M_k^{[.]}\right)d\zeta(\Theta_k)}\ra 1.
\]
For all $\omega\in\Omega$ there exists $\eta^*$ such that for all $n>\eta^*$ the posterior probability is
\[
p_{\mathcal N}\left(\mathfrak M_k^{[.]}\big|\mathbf y(\omega)\right)\geq 1-\lambda,\ \text{for all}\ k=1,2.
\]
Furthermore, the ratio of two posterior distributions is
\[
\frac{p\left(\mathfrak M_0^{[.]}\big|\mathbf y(\omega)\right)}{p\left(\mathfrak M_1^{[.]}\big|\mathbf y(\omega)\right)}=\frac{\pi\left(\Theta_0|\mathfrak M_0^{[.]}\right)}{\pi\left(\Theta_1|\mathfrak M_1^{[.]}\right)}\frac{\int_{\bm\Theta_k}\prod_{i=1}^nf^{[.]}\left(y_i(\omega)|\bm\Theta_1,\mathfrak M_1^{[.]}\right)\pi\left(\Theta_1|\mathfrak M_1^{[.]}\right)d\zeta(\Theta_1)}{\int_{\bm\Theta_k}\prod_{i=1}^nf^{[.]}\left(y_i(\omega)|\bm\Theta_0,\mathfrak M_0^{[.]}\right)\pi\left(\Theta_0|\mathfrak M_0^{[.]}\right)d\zeta(\Theta_0)}.
\]
For all $n>\eta^*$ and $\Theta_k\in\mathcal N$ yields
\begin{align*}
(1-\lambda)\left[\frac{\pi\left(\bar\Theta_0|\mathfrak M_0^{[.]}\right)}{\pi\left(\bar\Theta_1|\mathfrak M_1^{[.]}\right)}-\delta\right]& \left\{\frac{\int_{\mathcal N}\prod_{i=1}^nf^{[.]}\left(y_i(\omega)|\bm\Theta_1,\mathfrak M_1^{[.]}\right)\pi\left(\Theta_1|\mathfrak M_1^{[.]}\right)d\zeta(\Theta_1)}{\int_{\mathcal N}\prod_{i=1}^nf^{[.]}\left(y_i(\omega)|\bm\Theta_0,\mathfrak M_0^{[.]}\right)\pi\left(\Theta_0|\mathfrak M_0^{[.]}\right)d\zeta(\Theta_0)}\right\}\leq \frac{p\left(\mathfrak M_0^{[.]}\big|\mathbf y(\omega)\right)}{p\left(\mathfrak M_1^{[.]}\big|\mathbf y(\omega)\right)}\\
& \leq \frac{1}{(1-\lambda)}\left[\frac{\pi\left(\bar\Theta_0|\mathfrak M_0^{[.]}\right)}{\pi\left(\bar\Theta_1|\mathfrak M_1^{[.]}\right)}+\delta\right]\left\{ \frac{\int_{\mathcal N}\prod_{i=1}^nf^{[.]}\left(y_i(\omega)|\bm\Theta_1,\mathfrak M_1^{[.]}\right)\pi\left(\Theta_1|\mathfrak M_1^{[.]}\right)d\zeta(\Theta_1)}{\int_{\mathcal N}\prod_{i=1}^nf^{[.]}\left(y_i(\omega)|\bm\Theta_0,\mathfrak M_0^{[.]}\right)\pi\left(\Theta_0|\mathfrak M_0^{[.]}\right)d\zeta(\Theta_0)}\right\},
\end{align*}
and by the choice of $\mathcal N$,
\begin{align}\label{4}
\left[\pi\left(\bar\Theta_k|\mathfrak M_k^{[.]}\right)-\delta\right]&\int_{\mathcal N}\prod_{i=1}^nf^{[.]}\left(y_i(\omega)|\bm\Theta_0,\mathfrak M_0^{[.]}\right)d\zeta(\Theta_0)\leq \int_{\mathcal N}\prod_{i=1}^nf^{[.]}\left(y_i(\omega)|\bm\Theta_0,\mathfrak M_0^{[.]}\right)\pi\left(\Theta_0|\mathfrak M_0^{[.]}\right)d\zeta(\Theta_0)\notag\\
&\leq \left[\pi\left(\bar\Theta_k|\mathfrak M_k^{[.]}\right)+\delta\right]\int_{\mathcal N}\prod_{i=1}^nf^{[.]}\left(y_i(\omega)|\bm\Theta_0,\mathfrak M_0^{[.]}\right)d\zeta(\Theta_0).
\end{align}
For $\Theta_k\in\mathcal N$ the inequality \ref{4} yields
\begin{align*}
(1-\lambda)&\left[\frac{\pi\left(\bar\Theta_0|\mathfrak M_0^{[.]}\right)}{\pi\left(\bar\Theta_1|\mathfrak M_1^{[.]}\right)}-\delta\right]
\left[\frac{\pi\left(\bar\Theta_1|\mathfrak M_1^{[.]}\right)-\delta}{\pi\left(\bar\Theta_0|\mathfrak M_0^{[.]}\right)+\delta}\right]\leq \frac{p\left(\mathfrak M_0^{[.]}\big|\mathbf y(\omega)\right)}{p\left(\mathfrak M_1^{[.]}\big|\mathbf y(\omega)\right)}\\
&\leq\frac{1}{1-\lambda}\left[\frac{\pi\left(\bar\Theta_0|\mathfrak M_0^{[.]}\right)}{\pi\left(\bar\Theta_1|\mathfrak M_1^{[.]}\right)}+\delta\right]\left[\frac{\pi\left(\bar\Theta_1|\mathfrak M_1^{[.]}\right)+\delta}{\pi\left(\bar\Theta_0|\mathfrak M_0^{[.]}\right)-\delta}\right],
\end{align*}
so that for small values of $\delta$ and $\lambda$ we have
\[
\left|\frac{p\left(\mathfrak M_0^{[.]}\big|\mathbf y(\omega)\right)}{p\left(\mathfrak M_1^{[.]}\big|\mathbf y(\omega)\right)}\right|<\epsilon.
\]
Finally, for $n>\eta^*$,
\begin{align*}
\int_{\bm\Theta_k}&\left|p\left(\mathfrak M_0^{[.]}\big|\mathbf y(\omega)\right)-p\left(\mathfrak M_1^{[.]}\big|\mathbf y(\omega)\right)\right|d\zeta(\Theta_k)\\&\leq \int_{\mathcal N}p\left(\mathfrak M_1^{[.]}\big|\mathbf y(\omega)\right)\left|1-\frac{p\left(\mathfrak M_0^{[.]}\big|\mathbf y(\omega)\right)}{p\left(\mathfrak M_1^{[.]}\big|\mathbf y(\omega)\right)}\right|d\zeta(\Theta_k)+2\lambda\leq\epsilon(1-\lambda)+2\lambda.
\end{align*}
This completes the proof. $\qed$

\medskip

\subsection*{Proof of Lemma \ref{l00}}

\medskip

Consider two independent random variables $\varsigma_1$, $\varsigma_2$ with posterior probability distribution functions $p\left(\mathfrak M_0^{[.]}\big|\mathbf y\right)$, $p\left(\mathfrak M_1^{[.]}\big|\mathbf y\right)$ respectively. Then by  \cite{piterbarg1996},
\begin{align}\label{d0}
\mathbf p\left(\mathfrak M_k^{[.]}\bigg|\varsigma_1+\varsigma_2>\mathbf y,\varsigma_2\leq\mathbf y/4\right)&\leq\mathbf p\left(\mathfrak M_k^{[.]}\bigg|\varepsilon_1>3\mathbf y/4\right)=O\left(\mathbf y^{\varphi_0}\exp(-9\mathbf y^2/32)\right)\notag\\
\mathbf p\left(\mathfrak M_k^{[.]}\bigg|\varsigma_1+\varsigma_2>\mathbf y,\varsigma_2\leq 3\mathbf y/4\right)&\leq\mathbf p\left(\mathfrak M_k^{[.]}\bigg|\varepsilon_2>3\mathbf y/4\right)=O\left(\mathbf y^{\varphi_1}\exp(-9\mathbf y^2/32)\right),
\end{align}
where $\mathbf p=\left\{p\left(\mathfrak M_0^{[.]}\bigg|.\right),p\left(\mathfrak M_1^{[.]}\bigg|.\right)\right\}^T$ with $T$ represents the transposition of the matrix.

Now let us analyze the asymptotic properties of the finite integral
\begin{equation}\label{d1}
 I=\int_{\mathbf y/4}^{3\mathbf y/4}\left[1-p\left(\mathfrak M_0^{[.]}\bigg|\mathbf y-\mathbf z\right)\right]dp\left(\mathfrak M_1^{[.]}\bigg|\mathbf z\right).
\end{equation}
There exists $m>0$ such that $m\downarrow 0$. For a positive integer $\epsilon$ define $m_\epsilon=\epsilon m/\mathbf y$ so that $\mathbf y^2/(4m)$ is an integer. Therefore,
\begin{align}\label{d2}
I&\leq\sum_{\mathbf y/4\leq m_\epsilon\leq 3\mathbf y/4}\left[1-p\left(\mathfrak M_0^{[.]}\bigg|\mathbf y-m_\epsilon\right)\right]\left[p\left(\mathfrak M_1^{[.]}\bigg|m_\epsilon\right)-p\left(\mathfrak M_1^{[.]}\bigg|m_{\epsilon-1}\right)\right]\notag\\
I&\geq\sum_{\mathbf y/4\leq m_\epsilon\leq 3\mathbf y/4}\left[1-p\left(\mathfrak M_0^{[.]}\bigg|\mathbf y-m_{\epsilon-1}\right)\right]\left[p\left(\mathfrak M_1^{[.]}\bigg|m_\epsilon\right)-p\left(\mathfrak M_1^{[.]}\bigg|m_{\epsilon-1}\right)\right].
\end{align}
By condition 
\[
1-p\left(\mathfrak M_k^{[.]}\big|\mathbf y\right)=\Psi_k\ \mathbf y^{\varphi_k}\exp(-\mathbf y^2/2)(1+o(1)),\ \text{as $\mathbf y\ra\infty$,}
\]
there exist two monotonically decreasing functions $\eta_0(\mathbf y)\ra 0$, $\eta_1(\mathbf y)\ra0$ since $\mathbf y\ra \infty$ so that for all $\mathbf y>0$ we have,
\begin{equation*}
\Psi_k\mathbf y^{\varphi_k}\exp(\mathbf y^2/2)(1-\eta_k(\mathbf y))\leq 1-p\left(\mathfrak M_k^{[.]}\bigg|\mathbf y\right)
\leq \Psi_k\mathbf y^{\varphi_k}\exp(\mathbf y^2/2)(1+\eta_k(\mathbf y)),\ k=0,1.
\end{equation*}
Our main objective is to determine the estimate of the upper and the lower bounds of $p\left(\mathfrak M_1^{[.]}\bigg|m_\epsilon\right)-p\left(\mathfrak M_1^{[.]}\bigg|m_{\epsilon-1}\right)$ in condition \ref{d2}. The upper bound is
\begin{align*}
p&\left(\mathfrak M_1^{[.]}\bigg|m_\epsilon\right)-p\left(\mathfrak M_1^{[.]}\bigg|m_{\epsilon-1}\right)\\
&\leq \Psi_1 m_{\epsilon-1}^{\varphi_1}\exp\left(-\frac{1}{2}m_{\epsilon-1}^2\right)\left[1+\eta_1(m_{\epsilon-1})\right]-\Psi_1 m_{\epsilon}^{\varphi_1}\exp\left(-\frac{1}{2}m_{\epsilon}^2\right)\left[1-\eta_1(m_{\epsilon-1})\right]\\
&\leq \Psi_1 m_{\epsilon-1}^{\varphi_1}\exp\left(-\frac{1}{2}m_{\epsilon-1}^2\right)-\Psi_1 m_{\epsilon}^{\varphi_1}\exp\left(-\frac{1}{2}m_{\epsilon}^2\right)+2\Psi_1\eta_1(\mathbf y/4)m_{\epsilon-1}^{\varphi_1}\exp\left(-\frac{1}{2}m_{\epsilon-1}^2\right)\\
&=\Psi_1 m_{\epsilon}^{\varphi_1}\exp\left(-\frac{1}{2}m_{\epsilon}^2\right)\left[\left(\frac{\epsilon -1}{\epsilon}\right)\exp\left\{\frac{(2\epsilon-1)m^2}{2\mathbf y^2}\right\}-1\right]+2\Psi_1\eta_1(\mathbf y/4)m_{\epsilon-1}^{\varphi_1}\exp\left(-\frac{1}{2}m_{\epsilon-1}^2\right)\\
&\leq\Psi_1 m_\epsilon^{\phi_1}\exp\left(-\frac{1}{2}m_{\epsilon}^2\right)\left(\frac{\epsilon m^2}{\mathbf y^2}\right)\Upsilon\left(\frac{\epsilon m^2}{\mathbf y^2}\right)+2\Psi_1\eta_1(\mathbf y/4)m_{\epsilon-1}^{\varphi_1}\exp\left(-\frac{1}{2}m_{\epsilon-1}^2\right)\\
&\leq \Psi_1\Upsilon(3\epsilon/4)m_\epsilon^{\phi_1}\exp\left(-\frac{1}{2}m_{\epsilon}^2\right)\left(\frac{\epsilon m^2}{\mathbf y^2}\right)+2\Psi_1\eta_1(\mathbf y/4)m_{\epsilon-1}^{\varphi_1}\exp\left(-\frac{1}{2}m_{\epsilon-1}^2\right),
\end{align*}
where $\Upsilon(\mathbf y)=[\exp(\mathbf y)-1]/\mathbf y$.

The lower bound is,
\begin{align*}
p&\left(\mathfrak M_1^{[.]}\bigg|m_\epsilon\right)-p\left(\mathfrak M_1^{[.]}\bigg|m_{\epsilon-1}\right)\\
&\geq \left(\frac{\epsilon -1}{\epsilon}\right)^{\phi_1}\exp\left(\frac{-m^2}{2\mathbf y^2}\right)\Psi_1m_\epsilon^{\phi_1}\exp\left(-\frac{1}{2}m_{\epsilon}^2\right)\left(\frac{\epsilon m^2}{\mathbf y^2}\right)\Upsilon\left(\frac{\epsilon m^2}{\mathbf y^2}\right)-2\Psi_1\eta_1(\mathbf y/4)m_{\epsilon-1}^{\varphi_1}\exp\left(-\frac{1}{2}m_{\epsilon-1}^2\right)\\
&\geq\Upsilon(m/4)\left(1-\frac{4m}{\Upsilon\mathbf y^2}\right)\exp\left(\frac{-m^2}{2\mathbf y^2}\right)\Psi_1m_\epsilon^{\phi_1}\exp\left(-\frac{1}{2}m_{\epsilon}^2\right)\left(\frac{\epsilon m^2}{\mathbf y^2}\right)-2\Psi_1\eta_1(\mathbf y/4)m_{\epsilon-1}^{\varphi_1}\exp\left(-\frac{1}{2}m_{\epsilon-1}^2\right).
\end{align*}

Therefore,
\begin{align}\label{d3}
I&\leq \left(1+\eta_0(\mathbf y/4)\right)\Upsilon(3m/4)\Psi_0\Psi_1\sum_{\mathbf y/4\leq m_\epsilon\leq 3\mathbf y/4}(\mathbf y-m_\epsilon)^{\phi_0}\exp\left(-\frac{(\mathbf y-m_\epsilon)^2}{2}\right)m_\epsilon^{\phi_1}\exp\left(-\frac{m_{\epsilon-1}^2}{2}\right)\left(\frac{\epsilon m^2}{\mathbf y^2}\right)\notag\\
&+2\left(1+\eta_0(\mathbf y/4)\right)\eta_1(\mathbf y/4)\Psi_0\Psi_1\sum_{\mathbf y/4\leq m_\epsilon\leq 3\mathbf y/4}(\mathbf y-m_\epsilon)^{\phi_0}\exp\left(-\frac{(\mathbf y-m_\epsilon)^2}{2}\right)m_\epsilon^{\phi_1}\exp\left(-\frac{m_{\epsilon-1}^2}{2}\right),
\end{align}
and
\begin{align}\label{d4}
I&\geq \left(1+\eta_0(\mathbf y/4)\right)\Upsilon(m/4)\Psi_0\Psi_1\left(1-\frac{4m}{3\mathbf y^2}\right)^{\phi_1}\exp\left(-\frac{m}{2\mathbf y^2}\right)\notag\\
&\times\sum_{\mathbf y/4\leq m_\epsilon\leq 3\mathbf y/4}(\mathbf y-m_\epsilon)^{\phi_0}\exp\left(-\frac{(\mathbf y-m_\epsilon)^2}{2}\right)m_\epsilon^{\phi_1}\exp\left(-\frac{m_{\epsilon-1}^2}{2}\right)\left(\frac{\epsilon m^2}{\mathbf y^2}\right)\notag\\
&-2\left(1+\eta_0(\mathbf y/4)\right)\eta_1(\mathbf y/4)\Psi_0\Psi_1\sum_{\mathbf y/4\leq m_\epsilon\leq 3\mathbf y/4}(\mathbf y-m_\epsilon)^{\phi_0}\exp\left(-\frac{(\mathbf y-m_\epsilon)^2}{2}\right)m_\epsilon^{\phi_1}\exp\left(-\frac{m_{\epsilon-1}^2}{2}\right).
\end{align}
Define $m_\epsilon':=m_\epsilon/\mathbf y=\epsilon m/\mathbf y^2$. First sum of the right hand side of Equation (\ref{d3}) yields,
\begin{align*}
I'&=\mathbf y^{\phi_0+\phi_1+2}\exp(-\mathbf y^2/2)\sum_{1/4\leq m_\epsilon'\leq 3/4}(1-m_\epsilon')^{\phi_0}m_\epsilon'^{\phi_1}\exp\left(\mathbf y^2(m_\epsilon'-m_\epsilon'^2)m_\epsilon'\right)\left(\frac{m}{\mathbf y^2}\right)\\
&=\mathbf y^{\phi_0+\phi_1+2}\exp(-\mathbf y^2/2)\sum_{-1/4\leq m_\epsilon'\leq 1/4}\left[\frac{1}{2}-\left(m_\epsilon'-\frac{1}{2}\right)\right]^{\phi_0}\left[\frac{1}{2}+\left(m_\epsilon'-\frac{1}{2}\right)\right]^{\phi_1}\exp\left\{\frac{\mathbf y^2}{4}-\mathbf y^2\left(\frac{1}{2}-m_\epsilon'\right)^2\right\}m_\epsilon'\frac{m}{\mathbf y^2}\\
&\leq 2\mathbf y^{\phi_0+\phi_1+2}\exp(-\mathbf y^2/4)\sum_{0\leq m_\epsilon'-(1/2)\leq 1/4}\left[\frac{1}{2}-\left(m_\epsilon'-\frac{1}{2}\right)\right]^{\phi_0}\left[\frac{1}{2}+\left(m_\epsilon'-\frac{1}{2}\right)\right]^{\phi_1}\exp\left\{-\mathbf y^2\left(\frac{1}{2}-m_\epsilon'\right)^2\right\}m_\epsilon'\frac{m}{\mathbf y^2}\\
& \leq 2\mathbf y^{\phi_0+\phi_1+2}\exp(-\mathbf y^2/4)\frac{1}{\mathbf y}\int_0^{1/4}\left(\frac{1}{2}-u+\frac{m}{\mathbf y^2}\right)^{\phi_0}\left(\frac{1}{2}+u+\frac{m}{\mathbf y^2}\right)^{\phi_1}\exp(-\mathbf y^2u^2)du\\
&=2\int_0^\infty\exp(-v^2)dv 2^{-(\phi_0+\phi_1)}\mathbf y^{\phi_0+\phi_1+1}\exp(-\mathbf y^2/4)(1+o(1)),
\end{align*}
as $\mathbf y\ra\infty$. The last inequality is obtained by using the monotone convergence theorem. The estimate from below for the first sum on the right hand side of condition (\ref{d4}) becomes,
\begin{align*}
\hat I'&=\mathbf y^{\phi_0+\phi_1+2}\exp(-\mathbf y^2/2)\sum_{1/4\leq m_\epsilon'\leq 3/4}(1-m_\epsilon')^{\phi_0}m_\epsilon'^{\phi_1}\exp\left(\mathbf y^2(m_\epsilon'-m_\epsilon'^2-m/\mathbf y^2)'\right)m_\epsilon\left(\frac{m}{\mathbf y^2}\right)\\
&=\exp(-m)y^{\phi_0+\phi_1+2}\exp(-\mathbf y^2/2)\sum_{-1/4\leq m_\epsilon'-(1/2)\leq 1/4}\left[\frac{1}{2}-\left(m_{\epsilon-1}'-\frac{1}{2}\right)\right]^{\phi_0}\left[\frac{1}{2}+\left(m_\epsilon'-\frac{1}{2}\right)\right]^{\phi_1}\\
&\hspace{3cm}\times\exp\left\{\frac{\mathbf y^2}{4}-\mathbf y^2\left(\frac{1}{2}-m_\epsilon'\right)^2\right\}m_\epsilon'\frac{m}{\mathbf y^2}\\
&\geq 2\exp(-m)y^{\phi_0+\phi_1+2}\exp(-\mathbf y^2/4)\sum_{0\leq m_\epsilon'-(1/2)\leq 1/4}\left[\frac{1}{2}-\left(m_{\epsilon}'-\frac{1}{2}\right)-\frac{m}{\mathbf y^2}\right]^{\phi_0}\left[\frac{1}{2}+\left(m_{\epsilon}'-\frac{1}{2}\right)-\frac{m}{\mathbf y^2}\right]^{\phi_1}\\
&\hspace{3cm}\times \exp\left\{-\mathbf y^2\left(\frac{1}{2}-m_\epsilon'\right)^2\right\}m_{\epsilon-1}'\frac{m}{\mathbf y^2}\\
&\geq 2\exp(-m)y^{\phi_0+\phi_1+2}\exp(-\mathbf y^2/4)\int_0^{1/4}\left(\frac{1}{2}-u+\frac{m}{\mathbf y^2}\right)^{\phi_0}\left(\frac{1}{2}+u+\frac{m}{\mathbf y^2}\right)^{\phi_1}\exp(-\mathbf y^2u^2)du\\
&=2\int_0^\infty\exp(-v^2)dv 2^{-(\phi_0+\phi_1)}\mathbf y^{\phi_0+\phi_1+1}\exp(-\mathbf y^2/4)(1+o(1)), 
\end{align*}
as $\mathbf y\ra\infty$. Now consider the second sums $I"$ and $\hat I"$, on the right hand side of condition (\ref{d3}). For all $\epsilon$, $\epsilon$-th summand in those sums is determined by the corresponding summand in the first sum by multiplying by $\mathbf y^2/(\epsilon m^2)\leq 4/m$. Dividing left hand side and right hand sides of condition (\ref{d3}) and (\ref{d4}) by
\[
V(\mathbf y)=2^{-(\phi_0+\phi_1)}\sqrt{\pi}\Psi_0\Psi_1\mathbf y^{\phi_0+\phi_1+1}\exp(-\mathbf y^2/4)
\]
and letting $\mathbf y\ra\infty$ yields,
\[
\Upsilon(4m/3)\exp(-m)\leq\liminf_{\mathbf y\ra\infty}\frac{I}{V(\mathbf y)}\leq\limsup_{\mathbf y\ra\infty}\frac{I}{V(\mathbf y)}\leq 1.
\]
For an arbitrary $m$, and definition of $\Upsilon$, the above inequality shows the asymptotic behavior of I. Therefore, the statement of the lemma follows by condition (\ref{d0}). This completes the proof. $\qed$

\medskip

\subsection*{Proof of Proposition \ref{p0}}

\medskip

In order to prove this proposition we would go along the line of \cite{ghosh2010}. Since $\tau=\sqrt{n}(\Theta_k-\Theta_k^n)$, under model $\mathfrak M_k^{[.]}$ we can write,
\begin{align*}
p\left(\tau,\mathfrak M_k^{[.]}\big|\mathbf y\right)&=\frac{f^{[.]}\left(\mathbf y\big|\Theta_k^n+\frac{\tau}{\sqrt{n}}\right)\pi\left(\Theta_k+\frac{\tau}{\sqrt{n}}\big|\mathfrak M_k^{[.]}\right)}{\int_{\mathbb R}f^{[.]}\left(\mathbf y\big|\Theta_k^n+\frac{u}{\sqrt{n}}\right)\pi\left(\Theta_k+\frac{u}{\sqrt{n}}\big|\mathfrak M_k^{[.]}\right)du}\\
&=\frac{\pi\left(\Theta_k+\frac{\tau}{\sqrt{n}}\big|\mathfrak M_k^{[.]}\right)\exp\left\{f^{n[.]}\left(\mathbf y\big|\Theta_k^n+\frac{\tau}{\sqrt{n}},\mathfrak M_k^{[.]}\right)-f^{n[.]}\left(\mathbf y\big|\Theta_k^n,\mathfrak M_k^{[.]}\right)\right\}}{\int_{\mathbb R}\pi\left(\Theta_k+\frac{u}{\sqrt{n}}\big|\mathfrak M_k^{[.]}\right)\exp\left\{f^{n[.]}\left(\mathbf y\big|\Theta_k^n+\frac{u}{\sqrt{n}},\mathfrak M_k^{[.]}\right)-f^{n[.]}\left(\mathbf y\big|\Theta_k^n,\mathfrak M_k^{[.]}\right)\right\}du}.
\end{align*}
It is sufficient to show that 
\begin{multline}\label{5}
\int_{\mathbb R}\left|\frac{\pi\left(\Theta_k+\frac{\tau}{\sqrt{n}}\big|\mathfrak M_k^{[.]}\right)\exp\left\{f^{n[.]}\left(\mathbf y\big|\Theta_k^n+\frac{\tau}{\sqrt{n}},\mathfrak M_k^{[.]}\right)-f^{n[.]}\left(\mathbf y\big|\Theta_k^n,\mathfrak M_k^{[.]}\right)\right\}}{\int_{\mathbb R}\pi\left(\Theta_k+\frac{u}{\sqrt{n}}\big|\mathfrak M_k^{[.]}\right)\exp\left\{f^{n[.]}\left(\mathbf y\big|\Theta_k^n+\frac{u}{\sqrt{n}},\mathfrak M_k^{[.]}\right)-f^{n[.]}\left(\mathbf y\big|\Theta_k^n,\mathfrak M_k^{[.]}\right)\right\}du}\right.\\\left.-\sqrt{\frac{\mathcal I\left(\bar\Theta_k\right)}{2\pi}}\exp\left\{-\frac{1}{2}\tau^2\mathcal I\left(\bar\Theta_k\right)\right\}\right|d\tau\overset{\mathcal P_{\bar\Theta_k}}{\to}0,
\end{multline}
or,
\begin{multline}\label{6}
\int_{\mathbb R}\left|\pi\left(\Theta_k+\frac{u}{\sqrt{n}}\big|\mathfrak M_k^{[.]}\right)\exp\left\{f^{n[.]}\left(\mathbf y\big|\Theta_k^n+\frac{u}{\sqrt{n}},\mathfrak M_k^{[.]}\right)-f^{n[.]}\left(\mathbf y\big|\Theta_k^n,\mathfrak M_k^{[.]}\right)\right\}\right.\\
\left.-\pi\left(\bar\Theta_k\big|\mathfrak M_k^{[.]}\right)\exp\left\{-\frac{1}{2}u^2\mathcal I\left(\bar\Theta_k\right)\right\}\right|du\overset{\mathcal P_{\bar\Theta_k}}{\to}0.
\end{multline}
In order to understand conditions \ref{5} and \ref{6} define
\[
\Upsilon_n:=\int_{\mathbb R}\pi\left(\Theta_k+\frac{u}{\sqrt{n}}\big|\mathfrak M_k^{[.]}\right)\exp\left\{f^{n[.]}\left(\mathbf y\big|\Theta_k^n+\frac{u}{\sqrt{n}},\mathfrak M_k^{[.]}\right)-f^{n[.]}\left(\mathbf y\big|\Theta_k^n,\mathfrak M_k^{[.]}\right)\right\}du.
\]
Thus, expression in condition \ref{5} becomes
\begin{multline*}
\frac{1}{\Upsilon_n}\left[\int_{\mathbb R}\left|\pi\left(\Theta_k+\frac{\tau}{\sqrt{n}}\big|\mathfrak M_k^{[.]}\right)\exp\left\{f^{n[.]}\left(\mathbf y\big|\Theta_k^n+\frac{\tau}{\sqrt{n}},\mathfrak M_k^{[.]}\right)-f^{n[.]}\left(\mathbf y\big|\Theta_k^n,\mathfrak M_k^{[.]}\right)\right\}\right.\right.\\
\left.\left.-\Upsilon_n \sqrt{\frac{\mathcal I\left(\bar\Theta_k\right)}{2\pi}}\exp\left\{-\frac{1}{2}\tau^2\mathcal I\left(\bar\Theta_k\right)\right\}\right|d\tau\right]\overset{\mathcal P_{\bar\Theta_k}}{\to}0.
\end{multline*}
Let us denote two integrals as
\begin{align*}
I_0:=&\int_{\mathbb R}\left|\pi\left(\Theta_k+\frac{\tau}{\sqrt{n}}\big|\mathfrak M_k^{[.]}\right)\exp\left\{f^{n[.]}\left(\mathbf y\big|\Theta_k^n+\frac{\tau}{\sqrt{n}},\mathfrak M_k^{[.]}\right)-f^{n[.]}\left(\mathbf y\big|\Theta_k^n,\mathfrak M_k^{[.]}\right)\right\}-\pi\left(\bar\Theta_k\big|\mathfrak M_k^{[.]}\right)\exp\left\{-\frac{1}{2}\tau^2\mathcal I\left(\bar\Theta_k\right)\right\}\right|d\tau\\
I_1&:=\int_{\mathbb R}\left|\pi\left(\bar\Theta_k\big|\mathfrak M_k^{[.]}\right)\exp\left\{-\frac{1}{2}\tau^2\mathcal I\left(\bar\Theta_k\right)\right\}-\Upsilon_n\sqrt{\frac{\mathcal I\left(\bar\Theta_k\right)}{2\pi}}\exp\left\{-\frac{1}{2}\tau^2\mathcal I\left(\bar\Theta_k\right)\right\}\right|d\tau.
\end{align*}
 Since condition \ref{6} implies $\Upsilon_n\ra \pi\left(\bar\Theta_k\big|\mathfrak M_k^{[.]}\right)\sqrt{2\pi/I(\bar\Theta_k)}$, it is sufficient to show that integral inside the parenthesis converges to zero in probability and this term is less than $I_0+I_1$. Now by condition \ref{6} $I_0\ra0$ and the expression in $I_1$,
 \[
 I_1=\left|\pi\left(\bar\Theta_k\big|\mathfrak M_k^{[.]}\right)-\Upsilon_n\sqrt{\frac{\mathcal I\left(\bar\Theta_k\right)}{2\pi}}\right|\int_{\mathbb R}\exp\left\{-\frac{1}{2}\tau^2\mathcal I\left(\bar\Theta_k\right)\right\}d\tau\ra 0,
 \]
 as $\Upsilon_n\ra \pi\left(\bar\Theta_k\big|\mathfrak M_k^{[.]}\right)\sqrt{2\pi/I(\bar\Theta_k)}$. For further simplicity of this problem, define a function 
 \[
 g_n:=-\frac{1}{n}\sum_{i=1}^n f_2^{[.]}\left(y_i\big|\Theta_k^n,\mathfrak M_k^{[.]}\right)=-\frac{1}{n}f_2^{[.]}\left(y_i\big|\Theta_k^n,\mathfrak M_k^{[.]}\right).
 \]
 Clearly $g_n\ra I(\Theta_k)$ almost surely in probability $\mathcal P_{\Theta_k}$ as $n\ra\oo$. To check condition \ref{6} it is sufficient to show that
 \begin{multline}
 \int_{\mathbb R}\left|\pi\left(\Theta_k+\frac{u}{\sqrt{n}}\big|\mathfrak M_k^{[.]}\right)\exp\left\{f^{n[.]}\left(\mathbf y\big|\Theta_k^n+\frac{u}{\sqrt{n}},\mathfrak M_k^{[.]}\right)-f^{n[.]}\left(\mathbf y\big|\Theta_k^n,\mathfrak M_k^{[.]}\right)\right\}\right.\\
 \left.-\pi\left(\Theta_k^n\big|\mathfrak M_k^{[.]}\right)\exp\left\{-\frac{1}{2}u^2g_n\right\}\right|du\overset{\mathcal P_{\bar\Theta_k}}{\to}0.
 \end{multline}
 For any $\delta,\varkappa>0$ let us break $\mathbb R$ into three regions so that
 $\mathcal C_1=\{u:|u|<\varkappa\ln\sqrt{n}\}$, $\mathcal C_2=\{u:\varkappa\ln\sqrt{n}<|u|<\delta\sqrt{n}\}$ and $\mathcal C_3=\{u:|u|>\delta\sqrt{n}\}$. For the region $\mathcal C_3$,
 \begin{multline*}
 \int_{\mathcal C_3}\left|\pi\left(\Theta_k+\frac{u}{\sqrt{n}}\big|\mathfrak M_k^{[.]}\right)\exp\left\{f^{n[.]}\left(\mathbf y\big|\Theta_k^n+\frac{u}{\sqrt{n}},\mathfrak M_k^{[.]}\right)-f^{n[.]}\left(\mathbf y\big|\Theta_k^n,\mathfrak M_k^{[.]}\right)\right\}-\pi\left(\Theta_k^n\big|\mathfrak M_k^{[.]}\right)\exp\left\{-\frac{1}{2}u^2g_n\right\}\right|du\\
 \leq\int_{\mathcal C_3}\pi\left(\Theta_k+\frac{u}{\sqrt{n}}\big|\mathfrak M_k^{[.]}\right)\exp\left\{f^{n[.]}\left(\mathbf y\big|\Theta_k^n+\frac{u}{\sqrt{n}},\mathfrak M_k^{[.]}\right)-f^{n[.]}\left(\mathbf y\big|\Theta_k^n,\mathfrak M_k^{[.]}\right)\right\}du+\int_{\mathcal C_3}\pi\left(\Theta_k^n\big|\mathfrak M_k^{[.]}\right)\exp\left\{-\frac{1}{2}u^2g_n\right\}du.
 \end{multline*}
 By Assumption (v) in \ref{a0} the first integral goes to zero and by the tail estimate of a normal distribution the second integral converges to zero \citep{ghosh2010}. Since, $\Theta_k^n\ra\Theta_k$ for $n\ra\oo$, then a Taylor series expansion yields,
 \[
 f^{n[.]}\left(\mathbf y\big|\Theta_k^n+\frac{u}{\sqrt{n}},\mathfrak M_k^{[.]}\right)-f^{n[.]}\left(\mathbf y\big|\Theta_k^n,\mathfrak M_k^{[.]}\right)=\frac{u^2}{2n}f_2^{n[.]}\left(\mathbf y\big|\Theta_k^n,\mathfrak M_k^{[.]}\right)+\frac{1}{6}\left(\frac{u}{\sqrt{n}}\right)^3f_3^{n[.]}\left(\mathbf y\big|\Theta_k^*,\mathfrak M_k^{[.]}\right)=-\frac{u^2g_n}{2}+R_n,
 \]
 where $\Theta_k^*\in(\bar\Theta_k,\Theta_k^n)$. Now for region $\mathcal C_1$
 \begin{multline*}
 \int_{\mathcal C_1}\left|\pi\left(\Theta_k+\frac{u}{\sqrt{n}}\big|\mathfrak M_k^{[.]}\right)\exp\left\{-\frac{u^2g_n}{2}+R_n\right\}-\pi\left(\Theta_k^n\big|\mathfrak M_k^{[.]}\right)\exp\left\{-\frac{1}{2}u^2g_n\right\}\right|du\\
 \leq \int_{\mathcal C_1}\pi\left(\Theta_k+\frac{u}{\sqrt{n}}\big|\mathfrak M_k^{[.]}\right)\left|\exp\left\{-\frac{u^2g_n}{2}+R_n\right\}-\exp\left\{-\frac{1}{2}u^2g_n\right\}\right|du\\
 +\int_{\mathcal C_1}\left|\pi\left(\Theta_k+\frac{u}{\sqrt{n}}\big|\mathfrak M_k^{[.]}\right)-\pi\left(\Theta_k^n\big|\mathfrak M_k^{[.]}\right)\right|\exp\left\{-\frac{1}{2}u^2g_n\right\}du.
 \end{multline*}
 Since the prior density $\pi(.|.)$ is continuous at $\bar\Theta_k$, the second integral converges to zero a.s. in probability $\mathcal P_{\bar\Theta_k}$.  The first integral of the above expression is,
 \begin{equation}\label{7}
 \int_{\mathcal C_1}\pi\left(\Theta_k+\frac{u}{\sqrt{n}}\big|\mathfrak M_k^{[.]}\right)\left|\exp\{R_n\}-1\right|\exp\left\{-\frac{1}{2}u^2g_n\right\}du\leq\int_{\mathcal C_1}\pi\left(\Theta_k+\frac{u}{\sqrt{n}}\big|\mathfrak M_k^{[.]}\right)|R_n|\exp|R_n|\exp\left\{-\frac{1}{2}u^2g_n\right\}du.
 \end{equation}
 Since 
 \[
 \sup_{u\in\mathcal C_1} R_n=\sup_{u\in\mathcal C_1}\left(\frac{u}{\sqrt{n}}\right)^3f_3^{n[.]}\left(\mathbf y\big|\Theta_k^*,\mathfrak M_k^{[.]}\right)\leq\frac{\varkappa^3}{n}(\ln\sqrt{n})^3O_P(1)=o_P(1),
 \]
 the condition \ref{7} satisfies
 \begin{multline*}
 \int_{\mathcal C_1}\pi\left(\Theta_k+\frac{u}{\sqrt{n}}\big|\mathfrak M_k^{[.]}\right)|R_n|\exp|R_n|\exp\left\{-\frac{1}{2}u^2g_n\right\}du\\
 \leq \pi\left(\Theta_k+\frac{u}{\sqrt{n}}\big|\mathfrak M_k^{[.]}\right)\int_{\mathcal C_1}\exp\left\{-\frac{1}{2}u^2g_n\right\}|R_n|\exp|R_n|du=o_P(1).
 \end{multline*}
 Finally, for the region $\mathcal C_2$,
 \begin{multline*}
 \int_{\mathcal C_2}\left|\pi\left(\Theta_k+\frac{u}{\sqrt{n}}\big|\mathfrak M_k^{[.]}\right)\exp\left\{-\frac{u^2g_n}{2}+R_n\right\}-\pi\left(\Theta_k^n\big|\mathfrak M_k^{[.]}\right)\exp\left\{-\frac{1}{2}u^2g_n\right\}\right|du\\
 \leq\int_{\mathcal C_2}\pi\left(\Theta_k+\frac{u}{\sqrt{n}}\big|\mathfrak M_k^{[.]}\right)\exp\left\{-\frac{u^2g_n}{2}+R_n\right\}du+\int_{\mathcal C_2}\pi\left(\Theta_k^n\big|\mathfrak M_k^{[.]}\right)\exp\left\{-\frac{1}{2}u^2g_n\right\}du.
 \end{multline*}
 For a large constant $C^*\in(0,\oo)$ the second integral of the above inequality satisfies
 \begin{multline*}
 \int_{\mathcal C_2}\pi\left(\Theta_k^n\big|\mathfrak M_k^{[.]}\right)\exp\left\{-\frac{1}{2}u^2g_n\right\}du\leq2\pi\left(\Theta_k^n\big|\mathfrak M_k^{[.]}\right)\exp\left\{-\frac{1}{2}\varkappa g_n\ln\sqrt{n}\right\}[\delta\sqrt{n}-\varkappa\ln\sqrt{n}]\\
 \leq C^*\pi\left(\Theta_k^n\big|\mathfrak M_k^{[.]}\right)\frac{\sqrt{n}}{n^{\varkappa g_n/4}}\overset{\mathcal P_{\bar\Theta_k}}{\to}0.
 \end{multline*}
 Since $u\in\mathcal C_2$ and $\varkappa\ln\sqrt{n}<|u|<\delta\sqrt{n}$, first integral yields $|u|/\sqrt{n}<\delta$. Therefore, \[
 |R_n|=\frac{1}{6}\left(\frac{u}{\sqrt{n}}\right)^3f_3^{n[.]}\left(\mathbf y\big|\Theta_k^*,\mathfrak M_k^{[.]}\right)\leq \frac{\delta u^2}{6n}f_3^{n[.]}\left(\mathbf y\big|\Theta_k^*,\mathfrak M_k^{[.]}\right).
 \]
 Small values of $\delta>0$ ensures
 \begin{equation}\label{8}
 \mathcal P_{\bar\Theta_k}\left\{|R_n|<\frac{u^2}{4}g_n,\ \forall u\in\mathcal C_2\right\}>1-\epsilon,\ \text{for $n>\eta^*$},
 \end{equation}
 as $\sup_{\Theta_k^*\in(\bar\Theta_k-\delta,\bar\Theta_k+\delta)}(1/n)\left|f_3^{n[.]}\left(\mathbf y\big|\Theta_k^*,\mathfrak M_k^{[.]}\right)\right|=O_P(1)$. The condition \ref{8} can be written as,
 \begin{equation}\label{9}
 \mathcal P_{\bar\Theta_k}\left\{-\frac{u^2}{2}g_n+R_n<-\frac{u^2}{4}g_n,\ \forall u\in\mathcal C_2\right\}>1-\epsilon.
 \end{equation} 
 Therefore, with probability greater than $1-\epsilon$,
 \begin{equation*}
 \int_{\mathcal C_2}\pi\left(\Theta_k+\frac{u}{\sqrt{n}}\big|\mathfrak M_k^{[.]}\right)\exp\left\{-\frac{u^2g_n}{2}+R_n\right\}du\leq\sup_{\Theta_k\in\mathcal C_2}\pi\left(\Theta_k+\frac{u}{\sqrt{n}}\big|\mathfrak M_k^{[.]}\right)\int_{\mathcal C_2}\exp\left\{-\frac{u^2g_n}{2}+R_n\right\}du\ra 0,
 \end{equation*}
 as $n\ra\oo$. Now first choosing a $\delta$ to ensure condition \ref{8} and then by working with the $\delta$ in first and second steps  yields the final expression. This completes the proof. $\qed$
 
 \medskip
 
 \subsection*{Derivation of Equation \ref{14}}
 
 \medskip

Note that the probability mass function (pmf) of the zero-truncated negative binomial random variable $Y$ is 
\[
f_T^{NB}(y|\kappa)=\frac{\frac{\Gamma(y+\gamma)}{y!\Gamma(\gamma)}\left(1+\frac{\kappa}{\gamma}\right)^{-\gamma}\left(1+\frac{\gamma}{\kappa}\right)^{-y}}{1-\left(1+\frac{\kappa}{\gamma}\right)^{-\gamma}},\ \ y=1,2,....
\]
The expected value of $Y$ is
\[
\E(Y)=\frac{\gamma^2}{\kappa\left[1-\left(1+\frac{\kappa}{\gamma}\right)^{-\gamma}\right]}.
\]
Note that
\[
\ln\left[f_T^{NB}(y|\kappa)\right]=\ln\left[\frac{\Gamma(y+\gamma)}{y!\Gamma(\gamma)}\right]-y\ln\left(1+\frac{\gamma}{\kappa}\right)-\gamma\ln\left[1-\left(1+\frac{\kappa}{\gamma}\right)^{-\gamma}\right].
\]
Since
\[
\frac{\partial}{\partial \kappa}f_T^{NB}(y|\kappa)=\frac{\gamma y}{\kappa^2\left(1+\frac{\gamma}{\kappa}\right)}-\frac{1}{1+\frac{\kappa}{\gamma}}-\frac{\left(1+\frac{\kappa}{\gamma}\right)^{-(1+\gamma)}}{1-\left(1+\frac{\kappa}{\gamma}\right)^{-\gamma}},
\]
and
\begin{align*}
\frac{\partial^2}{\partial \kappa^2}f_T^{NB}(y|\kappa)&=\frac{\gamma^2 y}{\kappa^4\left(1+\frac{\gamma}{\kappa}\right)^2}+\frac{1}{\gamma \left(1+\frac{\kappa}{\gamma}\right)^2}-\frac{2\gamma y}{\kappa^3\left(1+\frac{\gamma}{\kappa}\right)}-\frac{\left(1+\frac{1}{\gamma}\right)\left(1+\frac{\kappa}{\gamma}\right)^{-(2+\gamma)}}{1-\left(1+\frac{\kappa}{\gamma}\right)^{-\gamma}}-\frac{\left(1+\frac{\kappa}{\gamma}\right)^{-(2+\gamma)}}{\left[1-\left(1+\frac{\kappa}{\gamma}\right)^{-\gamma}\right]^2},
\end{align*}
we have the Fisher information matrix as
\begin{align*}
I(\kappa)&=-\E\left[\frac{\partial^2}{\partial \kappa^2}f_T^{NB}(y|\kappa)\right]\\
&=\left(\frac{\gamma}{\kappa}\right)^3\frac{1}{(\gamma+\kappa)\left[1-\left(1+\frac{\kappa}{\gamma}\right)^{-\gamma}\right]}\left[2-\frac{1}{(\gamma+\kappa)\left[1-\left(1+\frac{\kappa}{\gamma}\right)^{-\gamma}\right]}\right]-\frac{\gamma}{(\gamma+\kappa)^2}\\
&\hspace{2cm}+\frac{\left(1+\frac{\kappa}{\gamma}\right)^{-(2+\gamma)}}{1-\left(1+\frac{\kappa}{\gamma}\right)^{-\gamma}}\left[\frac{1}{\gamma}(1+\gamma)+\frac{1}{1-\left(1+\frac{\kappa}{\gamma}\right)^{-\gamma}}\right]\\
&=\frac{\gamma}{\kappa(\gamma+\kappa)}\left\{\left(\frac{\gamma}{\kappa}\right)^2\frac{1}{1-\left(1+\frac{\kappa}{\gamma}\right)^{-\gamma}}\left[2-\frac{1}{(\gamma+\kappa)\left[1-\left(1+\frac{\kappa}{\gamma}\right)^{-\gamma}\right]}\right]-\frac{\kappa}{\gamma+\kappa}\right.\\
&\left.\hspace{2cm}+\left[\frac{\kappa(\gamma+\kappa)}{\gamma}\right]\frac{\left(1+\frac{\kappa}{\gamma}\right)^{-(2+\gamma)}}{1-\left(1+\frac{\kappa}{\gamma}\right)^{-\gamma}}\left[\frac{1}{\gamma}(1+\gamma)+\frac{1}{1-\left(1+\frac{\kappa}{\gamma}\right)^{-\gamma}}\right]\right\}.
\end{align*}
Therefore, the Jeffreys prior is readily available by taking a squared root on $I(\kappa)$.

\bibliographystyle{apalike}
\bibliography{bib}	
\end{document}